\documentclass[aps,amsmath,amssymb,twocolumn,floatfix,prl,footinbib,superscriptaddress,longbibliography]{revtex4-1}

\usepackage{graphicx}
\usepackage{epstopdf}

\usepackage[applemac]{inputenc}
\usepackage[T1]{fontenc}
\usepackage{lmodern}
\usepackage[english]{babel}
\usepackage{ae}
\usepackage{units}
\usepackage{color}
\usepackage{url}

\usepackage{amsmath,amssymb,natbib,bm}
\usepackage{psfrag}
\usepackage{amssymb}
\usepackage{amsthm}
\usepackage{mathrsfs}  

\usepackage{fixltx2e}

\usepackage{slashed}

\usepackage[americaninductors]{circuitikz}
\usepackage{tikz}
\usetikzlibrary{arrows}

\usepackage[colorlinks]{hyperref}
\hypersetup{%
	plainpages=true,
	breaklinks=true,
	hypertexnames=false,
	pageanchor=true,
	colorlinks=true,
	linkcolor={blue},
	citecolor={red},
	urlcolor={blue},
	anchorcolor={black}
}

\usepackage{mleftright} 

\newcommand{\be}{\begin{equation}}
\newcommand{\ee}{\end{equation}}
\newcommand{\bea}{\begin{eqnarray}}
\newcommand{\eea}{\end{eqnarray}}

\newcommand{\abssq}[1]{\mleft| #1 \mright|^2}

\renewcommand{\eqref}[1]{\mbox{Eq.~(\ref{#1})}}

\newcommand{\figpanel}[2]{Fig.~\hyperref[#1]{\ref*{#1}(#2)}}
\newcommand{\figpanels}[3]{Fig.~\hyperref[#1]{\ref*{#1}(#2)-(#3)}}
\newcommand{\figpanelNoPrefix}[2]{\hyperref[#1]{\ref*{#1}(#2)}}

\AtBeginDocument{%
    \newwrite\bibnotes
    \def\bibnotesext{Notes.bib}
    \immediate\openout\bibnotes=\jobname\bibnotesext
    \immediate\write\bibnotes{@CONTROL{REVTEX41Control}}
    \immediate\write\bibnotes{@CONTROL{%
    apsrev41Control,author="08",editor="1",pages="0",title="0",year="1"}}
     \if@filesw
     \immediate\write\@auxout{\string\citation{apsrev41Control}}%
    \fi
}%

\begin{document}
\title{Organic charged polaritons in the ultrastrong coupling regime}

\author{Mao Wang}
\affiliation{Department of Chemistry and Molecular Biology, University of Gothenburg, 412 96 Gothenburg, Sweden}

\author{Suman Mallick}
\affiliation{Department of Chemistry and Molecular Biology, University of Gothenburg, 412 96 Gothenburg, Sweden}

\author{Anton Frisk Kockum}
\affiliation{Department of Microtechnology and Nanoscience, Chalmers University of Technology, 412 96 Gothenburg, Sweden}

\author{Karl B\"orjesson}
\email{karl.borjesson@gu.se}
\affiliation{Department of Chemistry and Molecular Biology, University of Gothenburg, 412 96 Gothenburg, Sweden}

\date{\today}

\begin{abstract}

We embedded an all-hydrocarbon-based carbocation in a metallic microcavity that was tuned to resonance with an electronic transition of the carbocation. The measured Rabi splitting was \unit[41]{\%} of the excitation energy, putting the system well into the ultrastrong coupling regime. Importantly, due to the intrinsic charge on the carbocation, the polaritons that form carry a significant charge fraction (\unit[0.55]{$e_0$}) and a large charge-to-mass ratio ($\sim\,$\unit[2400]{$e_0 / m_0$}). Moreover, the ground state of the ultrastrongly coupled system is calculated to carry about one percent of one elementary charge, implying a potentially enhanced charge transport across the molecules. These unique properties of our system, together with its convenient preparation, provide a practical platform to study charged polaritons in the ultrastrong coupling regime.

\end{abstract}

\maketitle


\paragraph{Introduction.}

Exciton-polaritons are hybrid light-matter quasiparticles formed when the electronic excitation in ``matter'' is strongly coupled to a resonant electromagnetic field. To reach the strong coupling regime, the coupling strength (i.e., half the Rabi splitting $\hbar\Omega$), which quantifies the light-matter coupling strength, must be larger than the overall dissipation of the system. When the coupling strength is further increased and approaches a significant fraction ($\gtrsim$\,\unit[10]{\%}) of the bare energy of the excitation ($E_x$), the system enters the so-called ultrastrong coupling (USC) regime~\cite{Kockum2019, Ciuti2005, Forn-Diaz2019}. In this regime, not only the excited state of the system, but also the ground state acquires photonic and excitonic contributions~\cite{Kockum2019, Ciuti2005, Forn-Diaz2019, Ashhab2010}. This change is due to the anti-resonant contribution in the interaction, which is usually ignored in standard strongly coupled systems, and leads to a renormalization of the electromagnetic field~\cite{Yu2021, Ye2021, Wang2021}. Theoretical investigations show that the virtual photons in the ground state can be released by nonadiabatic manipulation of the coupling strength~\cite{Gunter2009, DeLiberato2007, Beaudoin2011, DeLiberato2009, Takashima2008, Werlang2008, Carusotto2012, Garziano2013, Shapiro2015}. The new hybrid ground state may have modified attributes including ground-state chemical reactivity~\cite{Herrera2016}, ground-state electroluminescence~\cite{Cirio2016}, and electrical conductivity and optoelectrical properties~\cite{Gambino2014, Askenazi2014}.

The USC regime has been reached in several systems, including inorganic quantum well intersubband microcavities~\cite{Anappara2009}, superconducting circuits~\cite{Niemczyk2010}, organic molecules in metallic microcavities~\cite{Schwartz2011}, plasmonic nanoparticles~\cite{Mueller2020}, etc. (a detailed list of systems can be found in Ref.~\cite{Kockum2019}). Most of these systems demand precise fabrication or cryogenic conditions, which is challenging. More diverse and readily prepared systems are therefore needed in order to explore the new phenomena predicted in the USC regime and to move towards practical applications~\cite{Sanvitto2016}.

In this Letter, we show how trions can be used to reach USC in a system that only requires relatively simple preparation. Trions are excitons containing an extra hole or electron. When trions are strongly coupled to cavity photons, the newly formed polaritons are expected to carry a fraction of a net charge~\cite{Rapaport2000, Rapaport2001}. Trions therefore enable direct manipulation of the polariton behavior with external electrical or magnetic fields. Furthermore, charged polaritons have a strong interaction among each other because of Coulomb interaction and therefore constitute a promising platform to observe nonlinear phenomena~\cite{Daskalakis2014}.

Trions are generally unstable and can easily decompose into an exciton and a free charge at room temperature. Charged polaritons were therefore first achieved in inorganic quantum wells inside a microcavity at liquid-helium temperatures~\cite{Rapaport2000, Rapaport2001}. Later they were realized by strongly coupling trions in electrically doped semiconducting single-walled carbon nanotubes to cavity photons~\cite{Mohl2018}. However, the energy of trions in both inorganic quantum wells and nanotubes lie close to that of the neutral exciton. The formed charged polaritons are therefore always strongly admixed with the neutral exciton, making it challenging to study their intrinsic properties. For molecular systems, charged polaritons were recently reported by strongly coupling the electronic transition of hole-doped organic semiconductors in planar microcavities~\cite{Cheng2018, Krainova2020}. The coupled transition arises from electronic excitation from the lower-lying orbitals to the partially vacant highest occupied molecular orbital of the hole-doped molecules. This method establishes a path to achieve charged polaritons in molecules and open the possibility to electrically tune polariton properties in organic semiconductors, although the doping ratio was not unity (\unit[30]{wt\%}), and the highest possible collective coupling strength was therefore not achieved.

Here we demonstrate room-temperature stable charged polaritons in the USC regime by strongly coupling the electronic transition of a carbocation to the optical field in a metallic Fabry-P\'erot cavity. The carbocation carries an intrinsic net charge without extrinsic doping and can be regarded as a ``\unit[100]{\%} doped'' system. The obtained Rabi splitting can reach as high as $\sim\,$\unit[0.50]{eV}, amounting to \unit[41]{\%} of the corresponding exciton energy (\unit[1.22]{eV}), which puts the system well into the USC regime. The charge-to-mass ratio of the lower polaritonic state is estimated to be three orders of magnitude higher than that of an electron. More importantly, the ground state of the system carries a net charge of $\unit[0.01]{e_0}$ attributed to the effect of the USC. These unique properties of our system, together with its simple preparation, provide a practical platform to study charged polaritons in the USC regime.


\paragraph{Photophysics of the carbocation.}

\begin{figure}[]
\includegraphics[width=\linewidth]{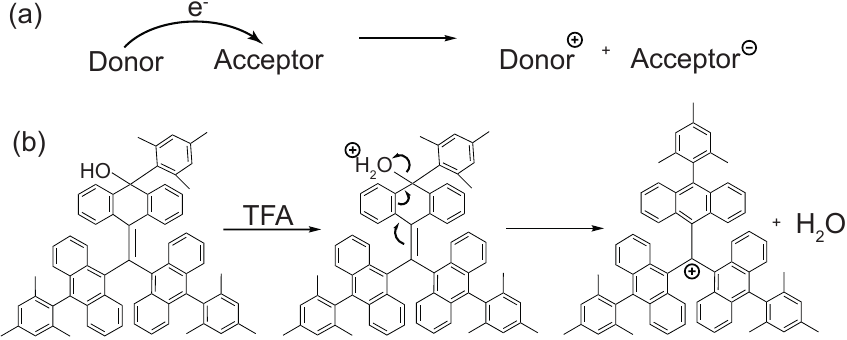}
\caption{
(a) Conventional doping using an electron transfer reaction from a donor to an acceptor molecule.
(b) Protonation using trifluoroacetic acid (TFA) followed by water elimination leading to the tris(10-mesitylanthr-9-yl)methyl cation (TAntM). 
\label{fig:DopingReaction}}
\end{figure}

In conventional chemical doping [\figpanel{fig:DopingReaction}{a}], an electron-transfer reaction occurs between the donor and acceptor molecules. In our system [\figpanel{fig:DopingReaction}{b}], the TAntM cation [tris(10-mesitylanthr-9-yl)methyl cation] was obtained by a two-step reaction: a proton transfer from trifluoroacetic acid (TFA) followed by an H$_2$O elimination reaction~\cite{Nishiuchi2018}. This two-step reaction is irreversible as H$_2$O is evaporated and the formed carbocation is, therefore, more stable than charged molecules formed by an electron-transfer reaction in the conventional doping method. Moreover, the formed carbocation, which usually is a chemically highly unstable species, is protected from nucleophilic attacks by the steric hindrance of the three surrounding bulky mesityl-substituted anthryl groups. These two effects result in unusually high stability for such an exotic molecule, allowing it to be handled under ambient conditions and solution processed into films.

\begin{figure}[]
\includegraphics[width=\linewidth]{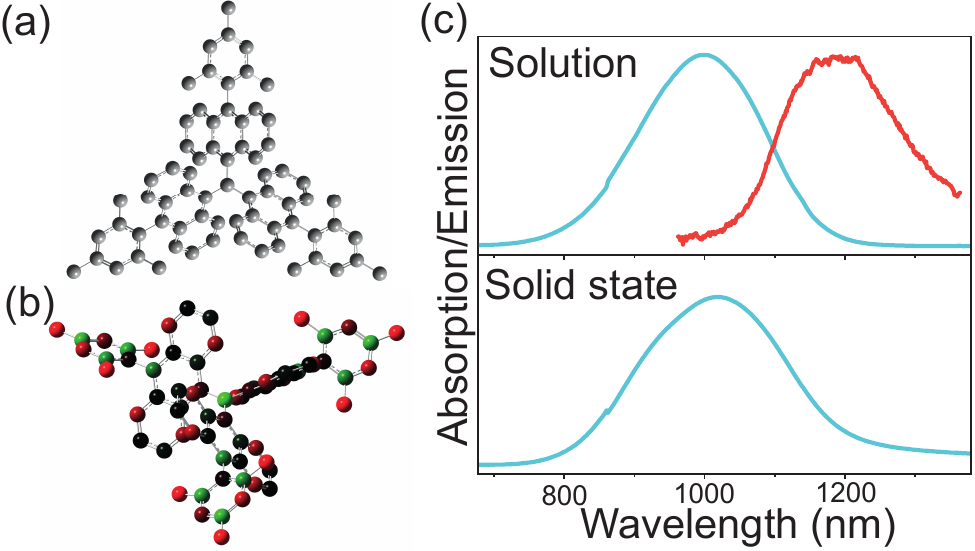}
\caption{
(a) Top view of the crystal structure of TAntM showing the 120-degree binding angle of the central carbon.
(b) Relative charge density map calculated by density functional theory showing atoms with partially positive charge as green and partially negative charge as red.
(c) Absorption and emission of TAntM in TFA solution and the absorption of a spin coated film of TAntM.
\label{fig:CrystalChargeAbsorptionEmission}}
\end{figure}

The positive charge on TAntM is central to its photophysical properties. Firstly, the single-crystal X-ray structure of the molecule shows that the central carbon forms in-plane bonds having binding angles of $120^\circ$ to the three bulky mesityl-substituted anthracene moieties~\cite{Nishiuchi2018}. Thus, the central carbon is confirmed to be sp$^2$-hybridized [\figpanel{fig:CrystalChargeAbsorptionEmission}{a}]. Based on the electrostatic potential surface of the molecule calculated using density functional theory [\figpanel{fig:CrystalChargeAbsorptionEmission}{b}], the central carbon carries a fraction of a positive charge. In addition, other carbons in the molecule have a fractional positive charge (although smaller in comparison). These carbons can all be predicted from resonance structures of the Lewis structure in \figpanel{fig:DopingReaction}{b}. The positive charge can therefore be considered somewhat delocalized over a considerable part of the molecule, increasing the chemical stability.

The characteristic feature of the absorption spectrum of the carbocation is an intense, broad band in the near-infrared regime [$\lambda_{\rm max} = \unit[990]{nm}$; \figpanel{fig:CrystalChargeAbsorptionEmission}{c}], with a peak molar absorption of $\sim\;$$\unit[3.7 \cdot 10^4]{M^{-1} cm^{-1}}$. This absorption band was assigned to the lowest-energy electronic transition ($S_1 \leftarrow S_0$), due to its mirror-image relationship to the emission spectrum. Further, time-dependent density-functional-theory calculations have previously predicted this transition to be, to a high degree, the result of a transition from the degenerate highest occupied molecular orbitals (HOMO and HOMO-1) to the lowest unoccupied molecular orbital (LUMO)~\cite{Nishiuchi2018}. The LUMO of the molecule contains the empty $p$ orbital of the positive central carbon, while the HOMOs were located on the surrounding anthracene moieties.

It is interesting to linger on the reason for the very low energy transition of the molecule, considering the small size of the aromatic network. The energy of the LUMO orbital is as low as $\unit[-5.81]{eV}$, a value comparable to the energy of the HOMO of many organic chromophores~\cite{Nishiuchi2018}. The reason for this low energy should be the empty $p$ orbital on the central carbon and the reason why a molecule having such a high electron affinity can be stable at all should be the bulkiness of the anthracene units. The anthracene bulk physically protects the $p$ orbital of the central carbon, resulting in increased stability kinetically, and provides an energetic penalty for the central carbon to adopt a tetrahedral geometry resulting in increased stability thermodynamically. This class of dye, having an empty carbon $p$ orbital, contains a positive charge on the LUMO and may furthermore even represent a direction towards closed-shell organic dyes having electronic transitions approaching vibrational energies. In summary, the molecular exciton formed can be regarded as a positive trion: one exciton bound to a positive charge. When this trion is strongly coupled to a cavity photon, the polaritons formed will therefore carry a fraction of the net charge.

Thin films of the carbocation were readily prepared by the spin-coating method. The envelope of absorption of the films was retained from the one in solution [\figpanel{fig:CrystalChargeAbsorptionEmission}{c}], suggesting that intermolecular interactions are weak. This point is important in the following discussion of the photophysics within the strong coupling regime, since it shows that each excited molecule can be regarded as an individual exciton. The peak absorption coefficient $\alpha$ of the film is $\unit[1.05 \cdot 10^5]{cm^{-1}}$. Based on the absorption cross-section $\sigma = \unit[6.14 \cdot 10^{-17}]{cm^2}$ derived from the molar absorption coefficient in solution, the charge density of the cation film is on the order of $\unit[10^{21}]{cm^{-3}}$, one order of magnitude higher than the density ($\unit[10^{20}]{cm^{-3}}$) in a previous report using extrinsic doping by co-evaporation of matrix and dopant molecules~\cite{Cheng2018}. As the coupling strength is proportional to the square root of the exciton density~\cite{Hertzog2019}, this high concentration indicates that the carbocation can be expected to achieve a much larger coupling strength and has the potential to reach the USC regime.


\paragraph{The ultrastrong coupling regime.}

\begin{figure}[]
\includegraphics[width=\linewidth]{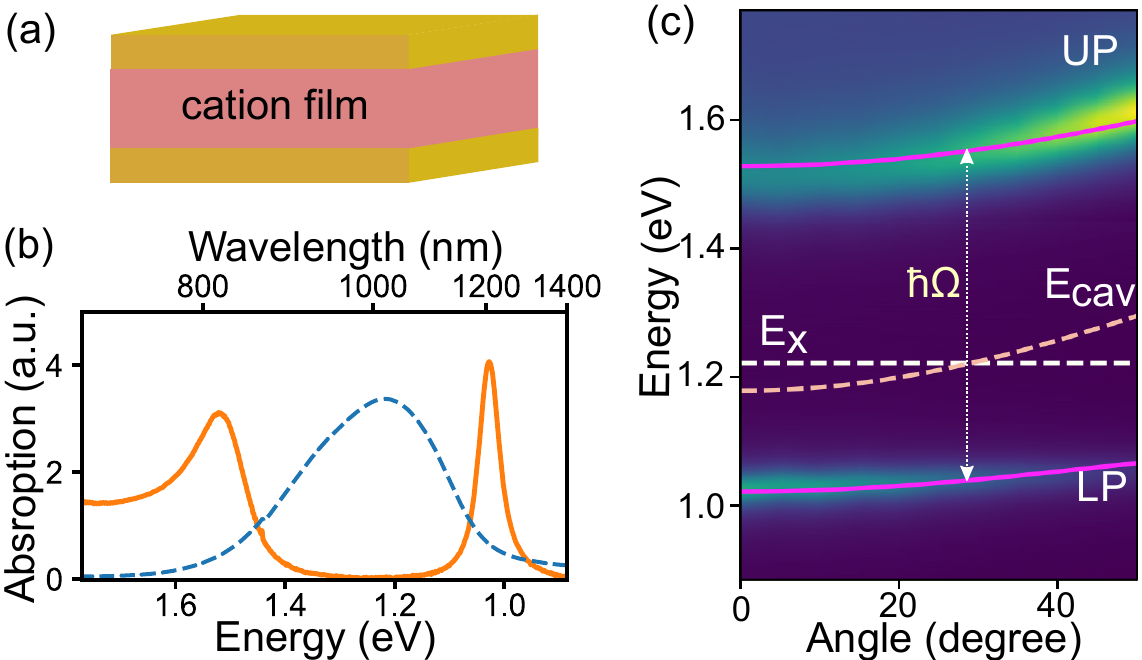}
\caption{
(a) The cavity structure: \unit[22]{nm} Au / TAntM film / \unit[22]{nm} Au.
(b) The absorption of a carbocation film (blue dashed line) outside and inside the cavity (orange solid line).
(c) The angular dispersion of the transmission spectra of a carbocation cavity. The purple solid lines are polaritonic branches obtained by fitting with the full Hopfield Hamiltonian, and the dashed white and orange lines are the exciton ($E_x$) and cavity ($E_{\rm cav}$) energies, respectively.
\label{fig:StructureSpectra}}
\end{figure}

To explore the (ultra-)strong coupling regime, a metallic Fabry-Perot cavity containing a carbocation film was fabricated. The active film was readily prepared by the spin-coating method (details in the Methods section in the Supplemental Material~\cite{SupMat}). The cavity structure is shown in \figpanel{fig:StructureSpectra}{a}. The active layer did not contain any polymer matrix, which otherwise is commonly used in the field. It therefore allowed the highest concentration possible and thus the largest collective light-matter coupling strength achievable using this molecule. A typical absorption spectrum of the cavity is shown in \figpanel{fig:StructureSpectra}{b}. The single absorption band of the bare film centered at \unit[1.22]{eV} is replaced by two polaritonic branches centered at 1.02 and \unit[1.52]{eV}.

On resonance, the energy difference between the two polaritonic branches is about \unit[0.5]{eV}, amounting to \unit[41]{\%} of the bare exciton energy (\unit[1.22]{eV}). This implies that the system is in the USC regime. In this regime, the rotating-wave approximation is not valid and the system should be treated using the full Hopfield Hamiltonian $H_{\rm Hop}$, where the anti-resonant and diamagnetic terms should be included (see Ref.~\cite{Kockum2019} for details).
%
%

The polaritonic energies $E$ are described analytically by the eigenvalues of $H_{\rm Hop}$, which can be obtained by solving the quadratic equation~\cite{Gambino2014}
\be
\mleft( E_{\rm cav}^2 - E^2 \mright) \mleft( E_x^2 - E^2 \mright) = \mleft( \hbar \Omega \mright)^2 E_{\rm cav}^2 ,
\label{eq:Energies}
\ee
where $E_x$ and $E_{\rm cav}$ are the energies of the bare exciton and cavity, respectively. Figure \figpanelNoPrefix{fig:StructureSpectra}{c} displays the angle-resolved transmission (TE mode). The polaritons show an angular dispersive behavior and a clear anticrossing feature is observed close to the exciton energy of the carbocation. By fitting the angular data to the results from the full Hopfield Hamiltonian, the obtained Rabi splitting is \unit[0.50]{eV}, which leads to a normalized coupling strength $\eta = \hbar \Omega / 2 E_x$ of 0.2. Thus, it confirms that the system is well into the USC regime. Notably, in the resonant case, the full treatment leads to an asymmetric anti-crossing in which the polaritonic energies follow~\cite{Gambino2014, Kena-Cohen2013} 
\be
E_{\rm UP/LP} = \sqrt{E_x^2 + \mleft( \frac{\hbar \Omega}{2} \mright)^2} \pm \frac{\hbar \Omega}{2} ,
\ee
due to a combined effect of the anti-resonant and diamagnetic terms. The energy of the lower polariton lies closer to the exciton energy at resonance ($E_x = E_{\rm cav}$) and become more matter-like compared to the case for the strong coupling regime [\figpanel{fig:StructureSpectra}{c}]. This anomalous effect is proportional to the normalized coupling strength $\eta$ and becomes significant in the USC regime~\cite{Ciuti2005}.

\begin{figure}[]
\includegraphics[width=\linewidth]{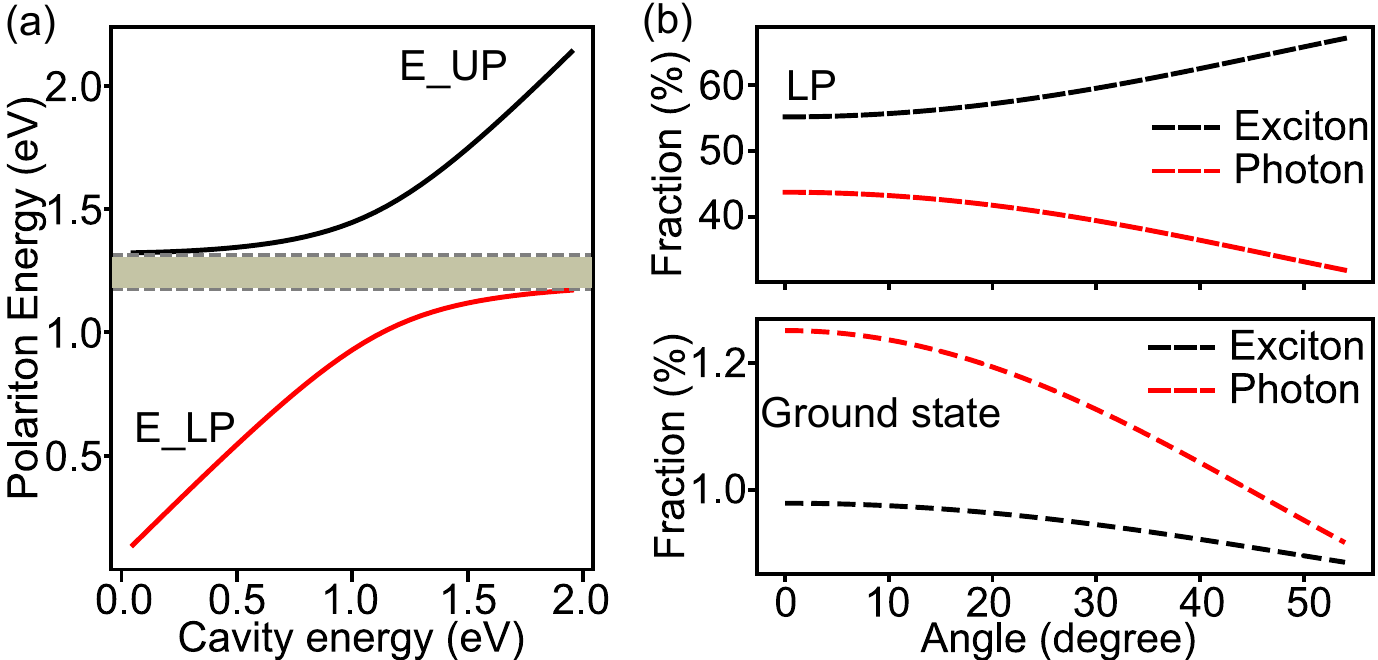}
\caption{
(a) Polariton energy as a function of cavity energy. The continuous line is the simulated polariton dispersion, using the parameters obtained from the fit of the angular dispersive transmission spectrum with \eqref{eq:Energies}. The shaded region indicates the polariton gap.
(b) The photon (exciton) ratio in the lower polariton (upper panel) and in the ground state (lower panel).
\label{fig:PolaritonGapExcitationFractions}}
\end{figure}

Another important feature of the USC regime is the appearance of a band gap of forbidden polariton energies, where no solution to \eqref{eq:Energies} will fall~\cite{Gambino2014, Todorov2010, Jouy2011, Delteil2012}. The forbidden frequencies correspond to destructive interference between the electromagnetic field radiated by the electronic oscillations and the bare microcavity photon field. As shown in \figpanel{fig:PolaritonGapExcitationFractions}{a}, the continuous lines represent the polaritonic branches obtained from the solution of \eqref{eq:Energies}. A polaritonic gap $E_g$ of \unit[130]{meV} is formed between the asymptotic lines of the upper and lower polaritonic dispersions [$E_g = \mleft( \hbar \Omega \mright)^2 / 2 E_x$]. The relative size of the gap ($E_g / E_x$) is proportional to $\eta^2$ and therefore only becomes significant in the USC regime. It can therefore also be regarded as a signature of the USC regime~\cite{Jouy2011} and has been observed previously~\cite{Gambino2014, Todorov2010, Jouy2011, Delteil2012}.


\paragraph{Charged polariton and ground-state modifications.}

Polaritons that are composed of charged excitons are expected to carry a net electric charge that is equal to the electron charge $e_0$ times the relative fraction of the charged exciton in the polariton mode. For example, the effective charge of the lower polariton is
%
$e_{\rm eff, LP} = e_0 \abssq{\alpha_{\rm LP, ex}}$,
%
where $\abssq{\alpha_{\rm LP, ex}}$ is the exciton fraction of the lower polariton. This effective value represents the probability of finding an electron bound to the polariton. The charge-to-mass ratio $e / m$ is a key parameter for the transport behavior of charged species under an electrical field. The polariton effective mass is the weighted harmonic mean of the mass of its exciton and photon components,
\be
\frac{1}{m_{\rm eff, LP}} = \frac{\abssq{\alpha_{\rm LP, ex}}}{m_{\rm ex}} + \frac{\abssq{\alpha_{\rm LP, ph}}}{m_{\rm ph}}
\label{eq:meffLP}
\ee
where $m_{\rm ex}$ and $m_{\rm ph}$ are the effective exciton and cavity-photon mass, respectively. As $m_{\rm ph} \ll m_{\rm ex}$, the effective polariton mass therefore mainly depends on the mass of the cavity photon, which is on the order of $10^{-4}$ times the bare electron mass $m_0$~\cite{Deng2010}. Due to this low effective mass, the charged polaritons therefore possess a much larger charge-to-mass ratio than typical charge carriers in organic semiconductors, enabling a potentially highly efficient charge transport.

In a previous report on a charged inorganic polariton~\cite{Rapaport2001}, the charge-to-mass ratio of polaritons was shown to be orders of magnitude larger than that of the free electron, and the enclosed theoretical calculations indicated a significantly higher drift velocity for the charged polaritons. Similar enhanced effects for charge transport was also discussed for trion-polaritons in semiconducting carbon nanotubes (CNTs)~\cite{Mohl2018}. The resulting charge-to-mass ratio in CNTs was about 200 times higher than that of an electron or hole. In our system, the net charge of the polariton is higher than that in inorganic or CNT systems, since the polaritonic states have no contribution from any neutral excitons. At zero probing angle ($k_\parallel = 0$), $e_{\rm eff, LP}$ is about $\unit[0.55]{e_0}$ and the effective charge-to-mass ratio $e_{\rm eff, LP} / m_{\rm eff, LP}$ of the charged polariton reaches a value of $\sim\,$\unit[2400]{$e_0 / m_0$} [$m_{\rm eff, LP}$ is calculated from \eqref{eq:meffLP}]. Moreover, the charge fraction of polaritons can be further increased by blue-detuning the cavity. The effective mass of electrons in organic semiconductors is generally large due to the strong localized charge and can reach as high as 25 times that of the bare electron mass~\cite{Coropceanu2007}. Thus, it is expected that charged polaritons will have several orders of magnitude higher charge-carrier mobility than typical charges in organic semiconductors. However, for organic electronics, the charge transport most often occurs in the ground state. Therefore, the formation of polaritonic states will have no obvious effect on the charge transport. This explains why only little attention has been devoted towards enhanced charge transport of organic semiconductors in the strong coupling regime~\cite{Nagarajan2020, Orgiu2015, Kang2021, Bhatt2021}.

In the USC regime, the treatment of the system with the full Hopfield Hamiltonian not only gives a more accurate description of the dispersion relation; it also indicates that the ground state is a squeezed vacuum state, which contains contributions from (virtual) photons and excitons. The effect of the photon contribution on the ground state was first discussed by Ciuti et al.~\cite{Ciuti2005}, who proposed the possibility of extracting the virtual photons by modulating the coupling strength, which is reminiscent of the dynamical Casimir effect~\cite{Moore1970}. Later, Cirio et al.~showed that electroluminescence can occur from the ground state in an ultrastrongly coupled system~\cite{Cirio2016, Cirio2019}.

Here we want to emphasize the effect of the exciton contribution on the polaritonic ground state. In our system, the ground state will carry a certain fraction of charge inherited from the exciton component. The photon and exciton components in the lower polariton and the ground state are shown in the upper and lower panels of \figpanel{fig:PolaritonGapExcitationFractions}{b}, respectively. In our red-detuned cavity, the energy of the cavity moves closer to the exciton energy as the angle increases and the LP becomes more exciton-like. However, both the excitonic and photonic contributions to the ground state decrease with increasing angle, a similar trend as in a previous report of an ultrastrongly coupled system by K\'ena-Cohen et al.~\cite{Kena-Cohen2013}. At zero degrees, the exciton contribution to the ground state of the system is about one percent. In this sense, the ground state will carry a fractional charge of $\unit[0.01]{e_0}$. This charge is delocalized over all molecules that contribute in the hybridization event that forms the polariton. Furthermore, it is up for discussion what the charge to mass ratio is for this fractional charge, although we expect it to be quite small. Of course, the ground state of the individual molecule is to the largest degree unperturbed, but the contribution of a delocalized fractional charge with a low charge to weight ratio implies a possibility of increasing the ground-state charge transport in organic materials. It is worth noting that this phenomenon only occurs when the charged polariton is in the USC regime, in which the ground state has a contribution from the charged excitons. Furthermore, how the fractional charge of the ground-state could be interpreted in any change of eventual ground-state mass, is open for discussion.


\paragraph{Conclusion.}

We have demonstrated the formation of charged polaritons in the USC regime at room temperature using a simple planar metallic microcavity encapsulating a carbocation film. The strategy using a transition from a lower lying orbital to HOMO allowed for a very low exciton energy, resulting in the normalized coupling strength reaching as high as \unit[20]{\%}, putting the system well into the USC regime. As a ``\unit[100]{\%} doped'' film, the formed polaritons can contain a significant charge fraction ($\unit[0.55]{e_0}$ for the lower polariton) and a large charge-to-mass ratio ($\sim\,$\unit[2400]{$e_0 / m_0$}). Moreover, there is a reflection of the excited state into the ground state in the USC regime, allowing the ground state of the system to carry about one percent of the elementary charge $e_0$, which is delocalized over the ensemble of molecules. It therefore implies the possibility of enhanced ground-state charge transport of organic materials in the USC regime through ground-state photoconductivity.


\begin{acknowledgments}

\paragraph{Acknowledgments.}

KB gratefully acknowledges financial support from the European Research Council (ERC-2017-StG-757733), the Swedish Research Council (2016-03354), and the Knut and Alice Wallenberg Foundation (KAW 2017.0192). AFK acknowledges support from the Swedish Research Council (grant number 2019-03696) and from the Knut and Alice Wallenberg Foundation through the Wallenberg Centre for Quantum Technology (WACQT).

\end{acknowledgments}


\bibliography{CationUSCRefs}

\begin{thebibliography}{46}%
\makeatletter
\providecommand \@ifxundefined [1]{%
 \@ifx{#1\undefined}
}%
\providecommand \@ifnum [1]{%
 \ifnum #1\expandafter \@firstoftwo
 \else \expandafter \@secondoftwo
 \fi
}%
\providecommand \@ifx [1]{%
 \ifx #1\expandafter \@firstoftwo
 \else \expandafter \@secondoftwo
 \fi
}%
\providecommand \natexlab [1]{#1}%
\providecommand \enquote  [1]{``#1''}%
\providecommand \bibnamefont  [1]{#1}%
\providecommand \bibfnamefont [1]{#1}%
\providecommand \citenamefont [1]{#1}%
\providecommand \href@noop [0]{\@secondoftwo}%
\providecommand \href [0]{\begingroup \@sanitize@url \@href}%
\providecommand \@href[1]{\@@startlink{#1}\@@href}%
\providecommand \@@href[1]{\endgroup#1\@@endlink}%
\providecommand \@sanitize@url [0]{\catcode `\\12\catcode `\$12\catcode
  `\&12\catcode `\#12\catcode `\^12\catcode `\_12\catcode `\%12\relax}%
\providecommand \@@startlink[1]{}%
\providecommand \@@endlink[0]{}%
\providecommand \url  [0]{\begingroup\@sanitize@url \@url }%
\providecommand \@url [1]{\endgroup\@href {#1}{\urlprefix }}%
\providecommand \urlprefix  [0]{URL }%
\providecommand \Eprint [0]{\href }%
\providecommand \doibase [0]{http://dx.doi.org/}%
\providecommand \selectlanguage [0]{\@gobble}%
\providecommand \bibinfo  [0]{\@secondoftwo}%
\providecommand \bibfield  [0]{\@secondoftwo}%
\providecommand \translation [1]{[#1]}%
\providecommand \BibitemOpen [0]{}%
\providecommand \bibitemStop [0]{}%
\providecommand \bibitemNoStop [0]{.\EOS\space}%
\providecommand \EOS [0]{\spacefactor3000\relax}%
\providecommand \BibitemShut  [1]{\csname bibitem#1\endcsname}%
\let\auto@bib@innerbib\@empty
\bibitem [{\citenamefont {Kockum}\ \emph {et~al.}(2019)\citenamefont {Kockum},
  \citenamefont {Miranowicz}, \citenamefont {{De Liberato}}, \citenamefont
  {Savasta},\ and\ \citenamefont {Nori}}]{Kockum2019}%
  \BibitemOpen
  \bibfield  {author} {\bibinfo {author} {\bibfnamefont {A.~F.}\ \bibnamefont
  {Kockum}}, \bibinfo {author} {\bibfnamefont {A.}~\bibnamefont {Miranowicz}},
  \bibinfo {author} {\bibfnamefont {S.}~\bibnamefont {{De Liberato}}}, \bibinfo
  {author} {\bibfnamefont {S.}~\bibnamefont {Savasta}}, \ and\ \bibinfo
  {author} {\bibfnamefont {F.}~\bibnamefont {Nori}},\ }\bibfield  {title}
  {\enquote {\bibinfo {title} {{Ultrastrong coupling between light and
  matter}},}\ }\href {\doibase 10.1038/s42254-018-0006-2} {\bibfield  {journal}
  {\bibinfo  {journal} {Nature Reviews Physics}\ }\textbf {\bibinfo {volume}
  {1}},\ \bibinfo {pages} {19} (\bibinfo {year} {2019})},\ \Eprint
  {http://arxiv.org/abs/1807.11636} {arXiv:1807.11636} \BibitemShut {NoStop}%
\bibitem [{\citenamefont {Ciuti}\ \emph {et~al.}(2005)\citenamefont {Ciuti},
  \citenamefont {Bastard},\ and\ \citenamefont {Carusotto}}]{Ciuti2005}%
  \BibitemOpen
  \bibfield  {author} {\bibinfo {author} {\bibfnamefont {C.}~\bibnamefont
  {Ciuti}}, \bibinfo {author} {\bibfnamefont {G.}~\bibnamefont {Bastard}}, \
  and\ \bibinfo {author} {\bibfnamefont {I.}~\bibnamefont {Carusotto}},\
  }\bibfield  {title} {\enquote {\bibinfo {title} {{Quantum vacuum properties
  of the intersubband cavity polariton field}},}\ }\href {\doibase
  10.1103/PhysRevB.72.115303} {\bibfield  {journal} {\bibinfo  {journal}
  {Physical Review B}\ }\textbf {\bibinfo {volume} {72}},\ \bibinfo {pages}
  {115303} (\bibinfo {year} {2005})},\ \Eprint {http://arxiv.org/abs/0504021}
  {arXiv:0504021 [cond-mat]} \BibitemShut {NoStop}%
\bibitem [{\citenamefont {Forn-D{\'{i}}az}\ \emph {et~al.}(2019)\citenamefont
  {Forn-D{\'{i}}az}, \citenamefont {Lamata}, \citenamefont {Rico},
  \citenamefont {Kono},\ and\ \citenamefont {Solano}}]{Forn-Diaz2019}%
  \BibitemOpen
  \bibfield  {author} {\bibinfo {author} {\bibfnamefont {P.}~\bibnamefont
  {Forn-D{\'{i}}az}}, \bibinfo {author} {\bibfnamefont {L.}~\bibnamefont
  {Lamata}}, \bibinfo {author} {\bibfnamefont {E.}~\bibnamefont {Rico}},
  \bibinfo {author} {\bibfnamefont {J.}~\bibnamefont {Kono}}, \ and\ \bibinfo
  {author} {\bibfnamefont {E.}~\bibnamefont {Solano}},\ }\bibfield  {title}
  {\enquote {\bibinfo {title} {{Ultrastrong coupling regimes of light-matter
  interaction}},}\ }\href {\doibase 10.1103/RevModPhys.91.025005} {\bibfield
  {journal} {\bibinfo  {journal} {Reviews of Modern Physics}\ }\textbf
  {\bibinfo {volume} {91}},\ \bibinfo {pages} {025005} (\bibinfo {year}
  {2019})},\ \Eprint {http://arxiv.org/abs/1804.09275} {arXiv:1804.09275}
  \BibitemShut {NoStop}%
\bibitem [{\citenamefont {Ashhab}\ and\ \citenamefont
  {Nori}(2010)}]{Ashhab2010}%
  \BibitemOpen
  \bibfield  {author} {\bibinfo {author} {\bibfnamefont {S.}~\bibnamefont
  {Ashhab}}\ and\ \bibinfo {author} {\bibfnamefont {F.}~\bibnamefont {Nori}},\
  }\bibfield  {title} {\enquote {\bibinfo {title} {{Qubit-oscillator systems in
  the ultrastrong-coupling regime and their potential for preparing
  nonclassical states}},}\ }\href {\doibase 10.1103/PhysRevA.81.042311}
  {\bibfield  {journal} {\bibinfo  {journal} {Physical Review A}\ }\textbf
  {\bibinfo {volume} {81}},\ \bibinfo {pages} {042311} (\bibinfo {year}
  {2010})},\ \Eprint {http://arxiv.org/abs/0912.4888} {arXiv:0912.4888}
  \BibitemShut {NoStop}%
\bibitem [{\citenamefont {Yu}\ \emph {et~al.}(2021)\citenamefont {Yu},
  \citenamefont {Mallick}, \citenamefont {Wang},\ and\ \citenamefont
  {B{\"{o}}rjesson}}]{Yu2021}%
  \BibitemOpen
  \bibfield  {author} {\bibinfo {author} {\bibfnamefont {Y.}~\bibnamefont
  {Yu}}, \bibinfo {author} {\bibfnamefont {S.}~\bibnamefont {Mallick}},
  \bibinfo {author} {\bibfnamefont {M.}~\bibnamefont {Wang}}, \ and\ \bibinfo
  {author} {\bibfnamefont {K.}~\bibnamefont {B{\"{o}}rjesson}},\ }\bibfield
  {title} {\enquote {\bibinfo {title} {{Barrier-free reverse-intersystem
  crossing in organic molecules by strong light-matter coupling}},}\ }\href
  {\doibase 10.1038/s41467-021-23481-6} {\bibfield  {journal} {\bibinfo
  {journal} {Nature Communications}\ }\textbf {\bibinfo {volume} {12}},\
  \bibinfo {pages} {3255} (\bibinfo {year} {2021})}\BibitemShut {NoStop}%
\bibitem [{\citenamefont {Ye}\ \emph {et~al.}(2021)\citenamefont {Ye},
  \citenamefont {Mallick}, \citenamefont {Hertzog}, \citenamefont
  {Kowalewski},\ and\ \citenamefont {B{\"{o}}rjesson}}]{Ye2021}%
  \BibitemOpen
  \bibfield  {author} {\bibinfo {author} {\bibfnamefont {C.}~\bibnamefont
  {Ye}}, \bibinfo {author} {\bibfnamefont {S.}~\bibnamefont {Mallick}},
  \bibinfo {author} {\bibfnamefont {M.}~\bibnamefont {Hertzog}}, \bibinfo
  {author} {\bibfnamefont {M.}~\bibnamefont {Kowalewski}}, \ and\ \bibinfo
  {author} {\bibfnamefont {K.}~\bibnamefont {B{\"{o}}rjesson}},\ }\bibfield
  {title} {\enquote {\bibinfo {title} {{Direct Transition from Triplet Excitons
  to Hybrid Light-Matter States via Triplet-Triplet Annihilation}},}\ }\href
  {\doibase 10.1021/jacs.1c02306} {\bibfield  {journal} {\bibinfo  {journal}
  {Journal of the American Chemical Society}\ }\textbf {\bibinfo {volume}
  {143}},\ \bibinfo {pages} {7501} (\bibinfo {year} {2021})}\BibitemShut
  {NoStop}%
\bibitem [{\citenamefont {Wang}\ \emph {et~al.}(2021)\citenamefont {Wang},
  \citenamefont {Hertzog},\ and\ \citenamefont {B{\"{o}}rjesson}}]{Wang2021}%
  \BibitemOpen
  \bibfield  {author} {\bibinfo {author} {\bibfnamefont {M.}~\bibnamefont
  {Wang}}, \bibinfo {author} {\bibfnamefont {M.}~\bibnamefont {Hertzog}}, \
  and\ \bibinfo {author} {\bibfnamefont {K.}~\bibnamefont {B{\"{o}}rjesson}},\
  }\bibfield  {title} {\enquote {\bibinfo {title} {{Polariton-assisted
  excitation energy channeling in organic heterojunctions}},}\ }\href {\doibase
  10.1038/s41467-021-22183-3} {\bibfield  {journal} {\bibinfo  {journal}
  {Nature Communications}\ }\textbf {\bibinfo {volume} {12}},\ \bibinfo {pages}
  {1874} (\bibinfo {year} {2021})}\BibitemShut {NoStop}%
\bibitem [{\citenamefont {G{\"{u}}nter}\ \emph {et~al.}(2009)\citenamefont
  {G{\"{u}}nter}, \citenamefont {Anappara}, \citenamefont {Hees}, \citenamefont
  {Sell}, \citenamefont {Biasiol}, \citenamefont {Sorba}, \citenamefont {{De
  Liberato}}, \citenamefont {Ciuti}, \citenamefont {Tredicucci}, \citenamefont
  {Leitenstorfer},\ and\ \citenamefont {Huber}}]{Gunter2009}%
  \BibitemOpen
  \bibfield  {author} {\bibinfo {author} {\bibfnamefont {G.}~\bibnamefont
  {G{\"{u}}nter}}, \bibinfo {author} {\bibfnamefont {A.~A.}\ \bibnamefont
  {Anappara}}, \bibinfo {author} {\bibfnamefont {J.}~\bibnamefont {Hees}},
  \bibinfo {author} {\bibfnamefont {A.}~\bibnamefont {Sell}}, \bibinfo {author}
  {\bibfnamefont {G.}~\bibnamefont {Biasiol}}, \bibinfo {author} {\bibfnamefont
  {L.}~\bibnamefont {Sorba}}, \bibinfo {author} {\bibfnamefont
  {S.}~\bibnamefont {{De Liberato}}}, \bibinfo {author} {\bibfnamefont
  {C.}~\bibnamefont {Ciuti}}, \bibinfo {author} {\bibfnamefont
  {A.}~\bibnamefont {Tredicucci}}, \bibinfo {author} {\bibfnamefont
  {A.}~\bibnamefont {Leitenstorfer}}, \ and\ \bibinfo {author} {\bibfnamefont
  {R.}~\bibnamefont {Huber}},\ }\bibfield  {title} {\enquote {\bibinfo {title}
  {{Sub-cycle switch-on of ultrastrong light--matter interaction}},}\ }\href
  {\doibase 10.1038/nature07838} {\bibfield  {journal} {\bibinfo  {journal}
  {Nature}\ }\textbf {\bibinfo {volume} {458}},\ \bibinfo {pages} {178}
  (\bibinfo {year} {2009})}\BibitemShut {NoStop}%
\bibitem [{\citenamefont {{De Liberato}}\ \emph {et~al.}(2007)\citenamefont
  {{De Liberato}}, \citenamefont {Ciuti},\ and\ \citenamefont
  {Carusotto}}]{DeLiberato2007}%
  \BibitemOpen
  \bibfield  {author} {\bibinfo {author} {\bibfnamefont {S.}~\bibnamefont {{De
  Liberato}}}, \bibinfo {author} {\bibfnamefont {C.}~\bibnamefont {Ciuti}}, \
  and\ \bibinfo {author} {\bibfnamefont {I.}~\bibnamefont {Carusotto}},\
  }\bibfield  {title} {\enquote {\bibinfo {title} {{Quantum Vacuum Radiation
  Spectra from a Semiconductor Microcavity with a Time-Modulated Vacuum Rabi
  Frequency}},}\ }\href {\doibase 10.1103/PhysRevLett.98.103602} {\bibfield
  {journal} {\bibinfo  {journal} {Physical Review Letters}\ }\textbf {\bibinfo
  {volume} {98}},\ \bibinfo {pages} {103602} (\bibinfo {year} {2007})},\
  \Eprint {http://arxiv.org/abs/0611282} {arXiv:0611282 [cond-mat]}
  \BibitemShut {NoStop}%
\bibitem [{\citenamefont {Beaudoin}\ \emph {et~al.}(2011)\citenamefont
  {Beaudoin}, \citenamefont {Gambetta},\ and\ \citenamefont
  {Blais}}]{Beaudoin2011}%
  \BibitemOpen
  \bibfield  {author} {\bibinfo {author} {\bibfnamefont {F.}~\bibnamefont
  {Beaudoin}}, \bibinfo {author} {\bibfnamefont {J.~M.}\ \bibnamefont
  {Gambetta}}, \ and\ \bibinfo {author} {\bibfnamefont {A.}~\bibnamefont
  {Blais}},\ }\bibfield  {title} {\enquote {\bibinfo {title} {{Dissipation and
  ultrastrong coupling in circuit QED}},}\ }\href {\doibase
  10.1103/PhysRevA.84.043832} {\bibfield  {journal} {\bibinfo  {journal}
  {Physical Review A}\ }\textbf {\bibinfo {volume} {84}},\ \bibinfo {pages}
  {043832} (\bibinfo {year} {2011})},\ \Eprint {http://arxiv.org/abs/1107.3990}
  {arXiv:1107.3990} \BibitemShut {NoStop}%
\bibitem [{\citenamefont {{De Liberato}}\ \emph {et~al.}(2009)\citenamefont
  {{De Liberato}}, \citenamefont {Gerace}, \citenamefont {Carusotto},\ and\
  \citenamefont {Ciuti}}]{DeLiberato2009}%
  \BibitemOpen
  \bibfield  {author} {\bibinfo {author} {\bibfnamefont {S.}~\bibnamefont {{De
  Liberato}}}, \bibinfo {author} {\bibfnamefont {D.}~\bibnamefont {Gerace}},
  \bibinfo {author} {\bibfnamefont {I.}~\bibnamefont {Carusotto}}, \ and\
  \bibinfo {author} {\bibfnamefont {C.}~\bibnamefont {Ciuti}},\ }\bibfield
  {title} {\enquote {\bibinfo {title} {{Extracavity quantum vacuum radiation
  from a single qubit}},}\ }\href {\doibase 10.1103/PhysRevA.80.053810}
  {\bibfield  {journal} {\bibinfo  {journal} {Physical Review A}\ }\textbf
  {\bibinfo {volume} {80}},\ \bibinfo {pages} {053810} (\bibinfo {year}
  {2009})},\ \Eprint {http://arxiv.org/abs/0906.2706} {arXiv:0906.2706}
  \BibitemShut {NoStop}%
\bibitem [{\citenamefont {Takashima}\ \emph {et~al.}(2008)\citenamefont
  {Takashima}, \citenamefont {Hatakenaka}, \citenamefont {Kurihara},\ and\
  \citenamefont {Zeilinger}}]{Takashima2008}%
  \BibitemOpen
  \bibfield  {author} {\bibinfo {author} {\bibfnamefont {K.}~\bibnamefont
  {Takashima}}, \bibinfo {author} {\bibfnamefont {N.}~\bibnamefont
  {Hatakenaka}}, \bibinfo {author} {\bibfnamefont {S.}~\bibnamefont
  {Kurihara}}, \ and\ \bibinfo {author} {\bibfnamefont {A.}~\bibnamefont
  {Zeilinger}},\ }\bibfield  {title} {\enquote {\bibinfo {title}
  {{Nonstationary boundary effect for a quantum flux in superconducting
  nanocircuits}},}\ }\href {\doibase 10.1088/1751-8113/41/16/164036} {\bibfield
   {journal} {\bibinfo  {journal} {Journal of Physics A: Mathematical and
  Theoretical}\ }\textbf {\bibinfo {volume} {41}},\ \bibinfo {pages} {164036}
  (\bibinfo {year} {2008})}\BibitemShut {NoStop}%
\bibitem [{\citenamefont {Werlang}\ \emph {et~al.}(2008)\citenamefont
  {Werlang}, \citenamefont {Dodonov}, \citenamefont {Duzzioni},\ and\
  \citenamefont {Villas-B{\^{o}}as}}]{Werlang2008}%
  \BibitemOpen
  \bibfield  {author} {\bibinfo {author} {\bibfnamefont {T.}~\bibnamefont
  {Werlang}}, \bibinfo {author} {\bibfnamefont {A.~V.}\ \bibnamefont
  {Dodonov}}, \bibinfo {author} {\bibfnamefont {E.~I.}\ \bibnamefont
  {Duzzioni}}, \ and\ \bibinfo {author} {\bibfnamefont {C.~J.}\ \bibnamefont
  {Villas-B{\^{o}}as}},\ }\bibfield  {title} {\enquote {\bibinfo {title} {{Rabi
  model beyond the rotating-wave approximation: Generation of photons from
  vacuum through decoherence}},}\ }\href {\doibase 10.1103/PhysRevA.78.053805}
  {\bibfield  {journal} {\bibinfo  {journal} {Physical Review A}\ }\textbf
  {\bibinfo {volume} {78}},\ \bibinfo {pages} {053805} (\bibinfo {year}
  {2008})},\ \Eprint {http://arxiv.org/abs/0806.3475} {arXiv:0806.3475}
  \BibitemShut {NoStop}%
\bibitem [{\citenamefont {Carusotto}\ \emph {et~al.}(2012)\citenamefont
  {Carusotto}, \citenamefont {{De Liberato}}, \citenamefont {Gerace},\ and\
  \citenamefont {Ciuti}}]{Carusotto2012}%
  \BibitemOpen
  \bibfield  {author} {\bibinfo {author} {\bibfnamefont {I.}~\bibnamefont
  {Carusotto}}, \bibinfo {author} {\bibfnamefont {S.}~\bibnamefont {{De
  Liberato}}}, \bibinfo {author} {\bibfnamefont {D.}~\bibnamefont {Gerace}}, \
  and\ \bibinfo {author} {\bibfnamefont {C.}~\bibnamefont {Ciuti}},\ }\bibfield
   {title} {\enquote {\bibinfo {title} {{Back-reaction effects of quantum
  vacuum in cavity quantum electrodynamics}},}\ }\href {\doibase
  10.1103/PhysRevA.85.023805} {\bibfield  {journal} {\bibinfo  {journal}
  {Physical Review A}\ }\textbf {\bibinfo {volume} {85}},\ \bibinfo {pages}
  {023805} (\bibinfo {year} {2012})}\BibitemShut {NoStop}%
\bibitem [{\citenamefont {Garziano}\ \emph {et~al.}(2013)\citenamefont
  {Garziano}, \citenamefont {Ridolfo}, \citenamefont {Stassi}, \citenamefont
  {{Di Stefano}},\ and\ \citenamefont {Savasta}}]{Garziano2013}%
  \BibitemOpen
  \bibfield  {author} {\bibinfo {author} {\bibfnamefont {L.}~\bibnamefont
  {Garziano}}, \bibinfo {author} {\bibfnamefont {A.}~\bibnamefont {Ridolfo}},
  \bibinfo {author} {\bibfnamefont {R.}~\bibnamefont {Stassi}}, \bibinfo
  {author} {\bibfnamefont {O.}~\bibnamefont {{Di Stefano}}}, \ and\ \bibinfo
  {author} {\bibfnamefont {S.}~\bibnamefont {Savasta}},\ }\bibfield  {title}
  {\enquote {\bibinfo {title} {{Switching on and off of ultrastrong
  light-matter interaction: Photon statistics of quantum vacuum radiation}},}\
  }\href {\doibase 10.1103/PhysRevA.88.063829} {\bibfield  {journal} {\bibinfo
  {journal} {Physical Review A}\ }\textbf {\bibinfo {volume} {88}},\ \bibinfo
  {pages} {063829} (\bibinfo {year} {2013})}\BibitemShut {NoStop}%
\bibitem [{\citenamefont {Shapiro}\ \emph {et~al.}(2015)\citenamefont
  {Shapiro}, \citenamefont {Zhukov}, \citenamefont {Pogosov},\ and\
  \citenamefont {Lozovik}}]{Shapiro2015}%
  \BibitemOpen
  \bibfield  {author} {\bibinfo {author} {\bibfnamefont {D.~S.}\ \bibnamefont
  {Shapiro}}, \bibinfo {author} {\bibfnamefont {A.~A.}\ \bibnamefont {Zhukov}},
  \bibinfo {author} {\bibfnamefont {W.~V.}\ \bibnamefont {Pogosov}}, \ and\
  \bibinfo {author} {\bibfnamefont {Y.~E.}\ \bibnamefont {Lozovik}},\
  }\bibfield  {title} {\enquote {\bibinfo {title} {{Dynamical Lamb effect in a
  tunable superconducting qubit-cavity system}},}\ }\href {\doibase
  10.1103/PhysRevA.91.063814} {\bibfield  {journal} {\bibinfo  {journal}
  {Physical Review A}\ }\textbf {\bibinfo {volume} {91}},\ \bibinfo {pages}
  {063814} (\bibinfo {year} {2015})},\ \Eprint
  {http://arxiv.org/abs/1503.01666} {arXiv:1503.01666} \BibitemShut {NoStop}%
\bibitem [{\citenamefont {Herrera}\ and\ \citenamefont
  {Spano}(2016)}]{Herrera2016}%
  \BibitemOpen
  \bibfield  {author} {\bibinfo {author} {\bibfnamefont {F.}~\bibnamefont
  {Herrera}}\ and\ \bibinfo {author} {\bibfnamefont {F.~C.}\ \bibnamefont
  {Spano}},\ }\bibfield  {title} {\enquote {\bibinfo {title}
  {{Cavity-Controlled Chemistry in Molecular Ensembles}},}\ }\href {\doibase
  10.1103/PhysRevLett.116.238301} {\bibfield  {journal} {\bibinfo  {journal}
  {Physical Review Letters}\ }\textbf {\bibinfo {volume} {116}},\ \bibinfo
  {pages} {238301} (\bibinfo {year} {2016})},\ \Eprint
  {http://arxiv.org/abs/1512.05017} {arXiv:1512.05017} \BibitemShut {NoStop}%
\bibitem [{\citenamefont {Cirio}\ \emph {et~al.}(2016)\citenamefont {Cirio},
  \citenamefont {{De Liberato}}, \citenamefont {Lambert},\ and\ \citenamefont
  {Nori}}]{Cirio2016}%
  \BibitemOpen
  \bibfield  {author} {\bibinfo {author} {\bibfnamefont {M.}~\bibnamefont
  {Cirio}}, \bibinfo {author} {\bibfnamefont {S.}~\bibnamefont {{De
  Liberato}}}, \bibinfo {author} {\bibfnamefont {N.}~\bibnamefont {Lambert}}, \
  and\ \bibinfo {author} {\bibfnamefont {F.}~\bibnamefont {Nori}},\ }\bibfield
  {title} {\enquote {\bibinfo {title} {{Ground State Electroluminescence}},}\
  }\href {\doibase 10.1103/PhysRevLett.116.113601} {\bibfield  {journal}
  {\bibinfo  {journal} {Physical Review Letters}\ }\textbf {\bibinfo {volume}
  {116}},\ \bibinfo {pages} {113601} (\bibinfo {year} {2016})},\ \Eprint
  {http://arxiv.org/abs/1508.05849} {arXiv:1508.05849} \BibitemShut {NoStop}%
\bibitem [{\citenamefont {Gambino}\ \emph {et~al.}(2014)\citenamefont
  {Gambino}, \citenamefont {Mazzeo}, \citenamefont {Genco}, \citenamefont {{Di
  Stefano}}, \citenamefont {Savasta}, \citenamefont {Patan{\`{e}}},
  \citenamefont {Ballarini}, \citenamefont {Mangione}, \citenamefont {Lerario},
  \citenamefont {Sanvitto},\ and\ \citenamefont {Gigli}}]{Gambino2014}%
  \BibitemOpen
  \bibfield  {author} {\bibinfo {author} {\bibfnamefont {S.}~\bibnamefont
  {Gambino}}, \bibinfo {author} {\bibfnamefont {M.}~\bibnamefont {Mazzeo}},
  \bibinfo {author} {\bibfnamefont {A.}~\bibnamefont {Genco}}, \bibinfo
  {author} {\bibfnamefont {O.}~\bibnamefont {{Di Stefano}}}, \bibinfo {author}
  {\bibfnamefont {S.}~\bibnamefont {Savasta}}, \bibinfo {author} {\bibfnamefont
  {S.}~\bibnamefont {Patan{\`{e}}}}, \bibinfo {author} {\bibfnamefont
  {D.}~\bibnamefont {Ballarini}}, \bibinfo {author} {\bibfnamefont
  {F.}~\bibnamefont {Mangione}}, \bibinfo {author} {\bibfnamefont
  {G.}~\bibnamefont {Lerario}}, \bibinfo {author} {\bibfnamefont
  {D.}~\bibnamefont {Sanvitto}}, \ and\ \bibinfo {author} {\bibfnamefont
  {G.}~\bibnamefont {Gigli}},\ }\bibfield  {title} {\enquote {\bibinfo {title}
  {{Exploring Light--Matter Interaction Phenomena under Ultrastrong Coupling
  Regime}},}\ }\href {\doibase 10.1021/ph500266d} {\bibfield  {journal}
  {\bibinfo  {journal} {ACS Photonics}\ }\textbf {\bibinfo {volume} {1}},\
  \bibinfo {pages} {1042} (\bibinfo {year} {2014})}\BibitemShut {NoStop}%
\bibitem [{\citenamefont {Askenazi}\ \emph {et~al.}(2014)\citenamefont
  {Askenazi}, \citenamefont {Vasanelli}, \citenamefont {Delteil}, \citenamefont
  {Todorov}, \citenamefont {Andreani}, \citenamefont {Beaudoin}, \citenamefont
  {Sagnes},\ and\ \citenamefont {Sirtori}}]{Askenazi2014}%
  \BibitemOpen
  \bibfield  {author} {\bibinfo {author} {\bibfnamefont {B.}~\bibnamefont
  {Askenazi}}, \bibinfo {author} {\bibfnamefont {A.}~\bibnamefont {Vasanelli}},
  \bibinfo {author} {\bibfnamefont {A.}~\bibnamefont {Delteil}}, \bibinfo
  {author} {\bibfnamefont {Y.}~\bibnamefont {Todorov}}, \bibinfo {author}
  {\bibfnamefont {L.~C.}\ \bibnamefont {Andreani}}, \bibinfo {author}
  {\bibfnamefont {G.}~\bibnamefont {Beaudoin}}, \bibinfo {author}
  {\bibfnamefont {I.}~\bibnamefont {Sagnes}}, \ and\ \bibinfo {author}
  {\bibfnamefont {C.}~\bibnamefont {Sirtori}},\ }\bibfield  {title} {\enquote
  {\bibinfo {title} {{Ultra-strong light-matter coupling for designer
  Reststrahlen band}},}\ }\href {\doibase 10.1088/1367-2630/16/4/043029}
  {\bibfield  {journal} {\bibinfo  {journal} {New Journal of Physics}\ }\textbf
  {\bibinfo {volume} {16}},\ \bibinfo {pages} {043029} (\bibinfo {year}
  {2014})}\BibitemShut {NoStop}%
\bibitem [{\citenamefont {Anappara}\ \emph {et~al.}(2009)\citenamefont
  {Anappara}, \citenamefont {{De Liberato}}, \citenamefont {Tredicucci},
  \citenamefont {Ciuti}, \citenamefont {Biasiol}, \citenamefont {Sorba},\ and\
  \citenamefont {Beltram}}]{Anappara2009}%
  \BibitemOpen
  \bibfield  {author} {\bibinfo {author} {\bibfnamefont {A.~A.}\ \bibnamefont
  {Anappara}}, \bibinfo {author} {\bibfnamefont {S.}~\bibnamefont {{De
  Liberato}}}, \bibinfo {author} {\bibfnamefont {A.}~\bibnamefont
  {Tredicucci}}, \bibinfo {author} {\bibfnamefont {C.}~\bibnamefont {Ciuti}},
  \bibinfo {author} {\bibfnamefont {G.}~\bibnamefont {Biasiol}}, \bibinfo
  {author} {\bibfnamefont {L.}~\bibnamefont {Sorba}}, \ and\ \bibinfo {author}
  {\bibfnamefont {F.}~\bibnamefont {Beltram}},\ }\bibfield  {title} {\enquote
  {\bibinfo {title} {{Signatures of the ultrastrong light-matter coupling
  regime}},}\ }\href {\doibase 10.1103/PhysRevB.79.201303} {\bibfield
  {journal} {\bibinfo  {journal} {Physical Review B}\ }\textbf {\bibinfo
  {volume} {79}},\ \bibinfo {pages} {201303} (\bibinfo {year} {2009})},\
  \Eprint {http://arxiv.org/abs/0808.3720} {arXiv:0808.3720} \BibitemShut
  {NoStop}%
\bibitem [{\citenamefont {Niemczyk}\ \emph {et~al.}(2010)\citenamefont
  {Niemczyk}, \citenamefont {Deppe}, \citenamefont {Huebl}, \citenamefont
  {Menzel}, \citenamefont {Hocke}, \citenamefont {Schwarz}, \citenamefont
  {Garcia-Ripoll}, \citenamefont {Zueco}, \citenamefont {H{\"{u}}mmer},
  \citenamefont {Solano}, \citenamefont {Marx},\ and\ \citenamefont
  {Gross}}]{Niemczyk2010}%
  \BibitemOpen
  \bibfield  {author} {\bibinfo {author} {\bibfnamefont {T.}~\bibnamefont
  {Niemczyk}}, \bibinfo {author} {\bibfnamefont {F.}~\bibnamefont {Deppe}},
  \bibinfo {author} {\bibfnamefont {H.}~\bibnamefont {Huebl}}, \bibinfo
  {author} {\bibfnamefont {E.~P.}\ \bibnamefont {Menzel}}, \bibinfo {author}
  {\bibfnamefont {F.}~\bibnamefont {Hocke}}, \bibinfo {author} {\bibfnamefont
  {M.~J.}\ \bibnamefont {Schwarz}}, \bibinfo {author} {\bibfnamefont {J.~J.}\
  \bibnamefont {Garcia-Ripoll}}, \bibinfo {author} {\bibfnamefont
  {D.}~\bibnamefont {Zueco}}, \bibinfo {author} {\bibfnamefont
  {T.}~\bibnamefont {H{\"{u}}mmer}}, \bibinfo {author} {\bibfnamefont
  {E.}~\bibnamefont {Solano}}, \bibinfo {author} {\bibfnamefont
  {A.}~\bibnamefont {Marx}}, \ and\ \bibinfo {author} {\bibfnamefont
  {R.}~\bibnamefont {Gross}},\ }\bibfield  {title} {\enquote {\bibinfo {title}
  {{Circuit quantum electrodynamics in the ultrastrong-coupling regime}},}\
  }\href {\doibase 10.1038/nphys1730} {\bibfield  {journal} {\bibinfo
  {journal} {Nature Physics}\ }\textbf {\bibinfo {volume} {6}},\ \bibinfo
  {pages} {772} (\bibinfo {year} {2010})},\ \Eprint
  {http://arxiv.org/abs/1003.2376} {arXiv:1003.2376} \BibitemShut {NoStop}%
\bibitem [{\citenamefont {Schwartz}\ \emph {et~al.}(2011)\citenamefont
  {Schwartz}, \citenamefont {Hutchison}, \citenamefont {Genet},\ and\
  \citenamefont {Ebbesen}}]{Schwartz2011}%
  \BibitemOpen
  \bibfield  {author} {\bibinfo {author} {\bibfnamefont {T.}~\bibnamefont
  {Schwartz}}, \bibinfo {author} {\bibfnamefont {J.~A.}\ \bibnamefont
  {Hutchison}}, \bibinfo {author} {\bibfnamefont {C.}~\bibnamefont {Genet}}, \
  and\ \bibinfo {author} {\bibfnamefont {T.~W.}\ \bibnamefont {Ebbesen}},\
  }\bibfield  {title} {\enquote {\bibinfo {title} {{Reversible Switching of
  Ultrastrong Light-Molecule Coupling}},}\ }\href {\doibase
  10.1103/PhysRevLett.106.196405} {\bibfield  {journal} {\bibinfo  {journal}
  {Physical Review Letters}\ }\textbf {\bibinfo {volume} {106}},\ \bibinfo
  {pages} {196405} (\bibinfo {year} {2011})}\BibitemShut {NoStop}%
\bibitem [{\citenamefont {Mueller}\ \emph {et~al.}(2020)\citenamefont
  {Mueller}, \citenamefont {Okamura}, \citenamefont {Vieira}, \citenamefont
  {Juergensen}, \citenamefont {Lange}, \citenamefont {Barros}, \citenamefont
  {Schulz},\ and\ \citenamefont {Reich}}]{Mueller2020}%
  \BibitemOpen
  \bibfield  {author} {\bibinfo {author} {\bibfnamefont {N.~S.}\ \bibnamefont
  {Mueller}}, \bibinfo {author} {\bibfnamefont {Y.}~\bibnamefont {Okamura}},
  \bibinfo {author} {\bibfnamefont {B.~G.~M.}\ \bibnamefont {Vieira}}, \bibinfo
  {author} {\bibfnamefont {S.}~\bibnamefont {Juergensen}}, \bibinfo {author}
  {\bibfnamefont {H.}~\bibnamefont {Lange}}, \bibinfo {author} {\bibfnamefont
  {E.~B.}\ \bibnamefont {Barros}}, \bibinfo {author} {\bibfnamefont
  {F.}~\bibnamefont {Schulz}}, \ and\ \bibinfo {author} {\bibfnamefont
  {S.}~\bibnamefont {Reich}},\ }\bibfield  {title} {\enquote {\bibinfo {title}
  {{Deep strong light-matter coupling in plasmonic nanoparticle crystals}},}\
  }\href {\doibase 10.1038/s41586-020-2508-1} {\bibfield  {journal} {\bibinfo
  {journal} {Nature}\ }\textbf {\bibinfo {volume} {583}},\ \bibinfo {pages}
  {780} (\bibinfo {year} {2020})}\BibitemShut {NoStop}%
\bibitem [{\citenamefont {Sanvitto}\ and\ \citenamefont
  {K{\'{e}}na-Cohen}(2016)}]{Sanvitto2016}%
  \BibitemOpen
  \bibfield  {author} {\bibinfo {author} {\bibfnamefont {D.}~\bibnamefont
  {Sanvitto}}\ and\ \bibinfo {author} {\bibfnamefont {S.}~\bibnamefont
  {K{\'{e}}na-Cohen}},\ }\bibfield  {title} {\enquote {\bibinfo {title} {{The
  road towards polaritonic devices}},}\ }\href {\doibase 10.1038/nmat4668}
  {\bibfield  {journal} {\bibinfo  {journal} {Nature Materials}\ }\textbf
  {\bibinfo {volume} {15}},\ \bibinfo {pages} {1061} (\bibinfo {year}
  {2016})}\BibitemShut {NoStop}%
\bibitem [{\citenamefont {Rapaport}\ \emph {et~al.}(2000)\citenamefont
  {Rapaport}, \citenamefont {Harel}, \citenamefont {Cohen}, \citenamefont
  {Ron}, \citenamefont {Linder},\ and\ \citenamefont
  {Pfeiffer}}]{Rapaport2000}%
  \BibitemOpen
  \bibfield  {author} {\bibinfo {author} {\bibfnamefont {R.}~\bibnamefont
  {Rapaport}}, \bibinfo {author} {\bibfnamefont {R.}~\bibnamefont {Harel}},
  \bibinfo {author} {\bibfnamefont {E.}~\bibnamefont {Cohen}}, \bibinfo
  {author} {\bibfnamefont {A.}~\bibnamefont {Ron}}, \bibinfo {author}
  {\bibfnamefont {E.}~\bibnamefont {Linder}}, \ and\ \bibinfo {author}
  {\bibfnamefont {L.~N.}\ \bibnamefont {Pfeiffer}},\ }\bibfield  {title}
  {\enquote {\bibinfo {title} {{Negatively Charged Quantum Well Polaritons in a
  GaAs/AlAs Microcavity: An Analog of Atoms in a Cavity}},}\ }\href {\doibase
  10.1103/PhysRevLett.84.1607} {\bibfield  {journal} {\bibinfo  {journal}
  {Physical Review Letters}\ }\textbf {\bibinfo {volume} {84}},\ \bibinfo
  {pages} {1607} (\bibinfo {year} {2000})}\BibitemShut {NoStop}%
\bibitem [{\citenamefont {Rapaport}\ \emph {et~al.}(2001)\citenamefont
  {Rapaport}, \citenamefont {Cohen}, \citenamefont {Ron}, \citenamefont
  {Linder},\ and\ \citenamefont {Pfeiffer}}]{Rapaport2001}%
  \BibitemOpen
  \bibfield  {author} {\bibinfo {author} {\bibfnamefont {R.}~\bibnamefont
  {Rapaport}}, \bibinfo {author} {\bibfnamefont {E.}~\bibnamefont {Cohen}},
  \bibinfo {author} {\bibfnamefont {A.}~\bibnamefont {Ron}}, \bibinfo {author}
  {\bibfnamefont {E.}~\bibnamefont {Linder}}, \ and\ \bibinfo {author}
  {\bibfnamefont {L.~N.}\ \bibnamefont {Pfeiffer}},\ }\bibfield  {title}
  {\enquote {\bibinfo {title} {{Negatively charged polaritons in a
  semiconductor microcavity}},}\ }\href {\doibase 10.1103/PhysRevB.63.235310}
  {\bibfield  {journal} {\bibinfo  {journal} {Physical Review B}\ }\textbf
  {\bibinfo {volume} {63}},\ \bibinfo {pages} {235310} (\bibinfo {year}
  {2001})}\BibitemShut {NoStop}%
\bibitem [{\citenamefont {Daskalakis}\ \emph {et~al.}(2014)\citenamefont
  {Daskalakis}, \citenamefont {Maier}, \citenamefont {Murray},\ and\
  \citenamefont {K{\'{e}}na-Cohen}}]{Daskalakis2014}%
  \BibitemOpen
  \bibfield  {author} {\bibinfo {author} {\bibfnamefont {K.~S.}\ \bibnamefont
  {Daskalakis}}, \bibinfo {author} {\bibfnamefont {S.~A.}\ \bibnamefont
  {Maier}}, \bibinfo {author} {\bibfnamefont {R.}~\bibnamefont {Murray}}, \
  and\ \bibinfo {author} {\bibfnamefont {S.}~\bibnamefont {K{\'{e}}na-Cohen}},\
  }\bibfield  {title} {\enquote {\bibinfo {title} {{Nonlinear interactions in
  an organic polariton condensate}},}\ }\href {\doibase 10.1038/nmat3874}
  {\bibfield  {journal} {\bibinfo  {journal} {Nature Materials}\ }\textbf
  {\bibinfo {volume} {13}},\ \bibinfo {pages} {271} (\bibinfo {year}
  {2014})}\BibitemShut {NoStop}%
\bibitem [{\citenamefont {M{\"{o}}hl}\ \emph {et~al.}(2018)\citenamefont
  {M{\"{o}}hl}, \citenamefont {Graf}, \citenamefont {Berger}, \citenamefont
  {L{\"{u}}ttgens}, \citenamefont {Zakharko}, \citenamefont {Lumsargis},
  \citenamefont {Gather},\ and\ \citenamefont {Zaumseil}}]{Mohl2018}%
  \BibitemOpen
  \bibfield  {author} {\bibinfo {author} {\bibfnamefont {C.}~\bibnamefont
  {M{\"{o}}hl}}, \bibinfo {author} {\bibfnamefont {A.}~\bibnamefont {Graf}},
  \bibinfo {author} {\bibfnamefont {F.~J.}\ \bibnamefont {Berger}}, \bibinfo
  {author} {\bibfnamefont {J.}~\bibnamefont {L{\"{u}}ttgens}}, \bibinfo
  {author} {\bibfnamefont {Y.}~\bibnamefont {Zakharko}}, \bibinfo {author}
  {\bibfnamefont {V.}~\bibnamefont {Lumsargis}}, \bibinfo {author}
  {\bibfnamefont {M.~C.}\ \bibnamefont {Gather}}, \ and\ \bibinfo {author}
  {\bibfnamefont {J.}~\bibnamefont {Zaumseil}},\ }\bibfield  {title} {\enquote
  {\bibinfo {title} {{Trion-Polariton Formation in Single-Walled Carbon
  Nanotube Microcavities}},}\ }\href {\doibase 10.1021/acsphotonics.7b01549}
  {\bibfield  {journal} {\bibinfo  {journal} {ACS Photonics}\ }\textbf
  {\bibinfo {volume} {5}},\ \bibinfo {pages} {2074} (\bibinfo {year}
  {2018})}\BibitemShut {NoStop}%
\bibitem [{\citenamefont {Cheng}\ \emph {et~al.}(2018)\citenamefont {Cheng},
  \citenamefont {Dhanker}, \citenamefont {Gray}, \citenamefont {Mukhopadhyay},
  \citenamefont {Kennehan}, \citenamefont {Asbury}, \citenamefont {Sokolov},\
  and\ \citenamefont {Giebink}}]{Cheng2018}%
  \BibitemOpen
  \bibfield  {author} {\bibinfo {author} {\bibfnamefont {C.-Y.}\ \bibnamefont
  {Cheng}}, \bibinfo {author} {\bibfnamefont {R.}~\bibnamefont {Dhanker}},
  \bibinfo {author} {\bibfnamefont {C.~L.}\ \bibnamefont {Gray}}, \bibinfo
  {author} {\bibfnamefont {S.}~\bibnamefont {Mukhopadhyay}}, \bibinfo {author}
  {\bibfnamefont {E.~R.}\ \bibnamefont {Kennehan}}, \bibinfo {author}
  {\bibfnamefont {J.~B.}\ \bibnamefont {Asbury}}, \bibinfo {author}
  {\bibfnamefont {A.}~\bibnamefont {Sokolov}}, \ and\ \bibinfo {author}
  {\bibfnamefont {N.~C.}\ \bibnamefont {Giebink}},\ }\bibfield  {title}
  {\enquote {\bibinfo {title} {{Charged Polaron Polaritons in an Organic
  Semiconductor Microcavity}},}\ }\href {\doibase
  10.1103/PhysRevLett.120.017402} {\bibfield  {journal} {\bibinfo  {journal}
  {Physical Review Letters}\ }\textbf {\bibinfo {volume} {120}},\ \bibinfo
  {pages} {017402} (\bibinfo {year} {2018})}\BibitemShut {NoStop}%
\bibitem [{\citenamefont {Krainova}\ \emph {et~al.}(2020)\citenamefont
  {Krainova}, \citenamefont {Grede}, \citenamefont {Tsokkou}, \citenamefont
  {Banerji},\ and\ \citenamefont {Giebink}}]{Krainova2020}%
  \BibitemOpen
  \bibfield  {author} {\bibinfo {author} {\bibfnamefont {N.}~\bibnamefont
  {Krainova}}, \bibinfo {author} {\bibfnamefont {A.~J.}\ \bibnamefont {Grede}},
  \bibinfo {author} {\bibfnamefont {D.}~\bibnamefont {Tsokkou}}, \bibinfo
  {author} {\bibfnamefont {N.}~\bibnamefont {Banerji}}, \ and\ \bibinfo
  {author} {\bibfnamefont {N.~C.}\ \bibnamefont {Giebink}},\ }\bibfield
  {title} {\enquote {\bibinfo {title} {{Polaron Photoconductivity in the Weak
  and Strong Light-Matter Coupling Regime}},}\ }\href {\doibase
  10.1103/PhysRevLett.124.177401} {\bibfield  {journal} {\bibinfo  {journal}
  {Physical Review Letters}\ }\textbf {\bibinfo {volume} {124}},\ \bibinfo
  {pages} {177401} (\bibinfo {year} {2020})}\BibitemShut {NoStop}%
\bibitem [{\citenamefont {Nishiuchi}\ \emph {et~al.}(2018)\citenamefont
  {Nishiuchi}, \citenamefont {Aibara},\ and\ \citenamefont
  {Kubo}}]{Nishiuchi2018}%
  \BibitemOpen
  \bibfield  {author} {\bibinfo {author} {\bibfnamefont {T.}~\bibnamefont
  {Nishiuchi}}, \bibinfo {author} {\bibfnamefont {S.}~\bibnamefont {Aibara}}, \
  and\ \bibinfo {author} {\bibfnamefont {T.}~\bibnamefont {Kubo}},\ }\bibfield
  {title} {\enquote {\bibinfo {title} {{Synthesis and Properties of a Highly
  Congested Tri(9-anthryl)methyl Radical}},}\ }\href {\doibase
  10.1002/anie.201811314} {\bibfield  {journal} {\bibinfo  {journal}
  {Angewandte Chemie International Edition}\ }\textbf {\bibinfo {volume}
  {57}},\ \bibinfo {pages} {16516} (\bibinfo {year} {2018})}\BibitemShut
  {NoStop}%
\bibitem [{\citenamefont {Hertzog}\ \emph {et~al.}(2019)\citenamefont
  {Hertzog}, \citenamefont {Wang}, \citenamefont {Mony},\ and\ \citenamefont
  {B{\"{o}}rjesson}}]{Hertzog2019}%
  \BibitemOpen
  \bibfield  {author} {\bibinfo {author} {\bibfnamefont {M.}~\bibnamefont
  {Hertzog}}, \bibinfo {author} {\bibfnamefont {M.}~\bibnamefont {Wang}},
  \bibinfo {author} {\bibfnamefont {J.}~\bibnamefont {Mony}}, \ and\ \bibinfo
  {author} {\bibfnamefont {K.}~\bibnamefont {B{\"{o}}rjesson}},\ }\bibfield
  {title} {\enquote {\bibinfo {title} {{Strong light-matter interactions: a new
  direction within chemistry}},}\ }\href {\doibase 10.1039/C8CS00193F}
  {\bibfield  {journal} {\bibinfo  {journal} {Chemical Society Reviews}\
  }\textbf {\bibinfo {volume} {48}},\ \bibinfo {pages} {937} (\bibinfo {year}
  {2019})}\BibitemShut {NoStop}%
\bibitem [{Sup()}]{SupMat}%
  \BibitemOpen
  \href@noop {} {}\bibinfo {note} {See the Supplemental Material at
  XXX.}\BibitemShut {Stop}%
\bibitem [{\citenamefont {K{\'{e}}na-Cohen}\ \emph {et~al.}(2013)\citenamefont
  {K{\'{e}}na-Cohen}, \citenamefont {Maier},\ and\ \citenamefont
  {Bradley}}]{Kena-Cohen2013}%
  \BibitemOpen
  \bibfield  {author} {\bibinfo {author} {\bibfnamefont {S.}~\bibnamefont
  {K{\'{e}}na-Cohen}}, \bibinfo {author} {\bibfnamefont {S.~A.}\ \bibnamefont
  {Maier}}, \ and\ \bibinfo {author} {\bibfnamefont {D.~D.~C.}\ \bibnamefont
  {Bradley}},\ }\bibfield  {title} {\enquote {\bibinfo {title} {{Ultrastrongly
  Coupled Exciton-Polaritons in Metal-Clad Organic Semiconductor
  Microcavities}},}\ }\href {\doibase 10.1002/adom.201300256} {\bibfield
  {journal} {\bibinfo  {journal} {Advanced Optical Materials}\ }\textbf
  {\bibinfo {volume} {1}},\ \bibinfo {pages} {827} (\bibinfo {year}
  {2013})}\BibitemShut {NoStop}%
\bibitem [{\citenamefont {Todorov}\ \emph {et~al.}(2010)\citenamefont
  {Todorov}, \citenamefont {Andrews}, \citenamefont {Colombelli}, \citenamefont
  {{De Liberato}}, \citenamefont {Ciuti}, \citenamefont {Klang}, \citenamefont
  {Strasser},\ and\ \citenamefont {Sirtori}}]{Todorov2010}%
  \BibitemOpen
  \bibfield  {author} {\bibinfo {author} {\bibfnamefont {Y.}~\bibnamefont
  {Todorov}}, \bibinfo {author} {\bibfnamefont {A.~M.}\ \bibnamefont
  {Andrews}}, \bibinfo {author} {\bibfnamefont {R.}~\bibnamefont {Colombelli}},
  \bibinfo {author} {\bibfnamefont {S.}~\bibnamefont {{De Liberato}}}, \bibinfo
  {author} {\bibfnamefont {C.}~\bibnamefont {Ciuti}}, \bibinfo {author}
  {\bibfnamefont {P.}~\bibnamefont {Klang}}, \bibinfo {author} {\bibfnamefont
  {G.}~\bibnamefont {Strasser}}, \ and\ \bibinfo {author} {\bibfnamefont
  {C.}~\bibnamefont {Sirtori}},\ }\bibfield  {title} {\enquote {\bibinfo
  {title} {{Ultrastrong Light-Matter Coupling Regime with Polariton Dots}},}\
  }\href {\doibase 10.1103/PhysRevLett.105.196402} {\bibfield  {journal}
  {\bibinfo  {journal} {Physical Review Letters}\ }\textbf {\bibinfo {volume}
  {105}},\ \bibinfo {pages} {196402} (\bibinfo {year} {2010})},\ \Eprint
  {http://arxiv.org/abs/1301.1297} {arXiv:1301.1297} \BibitemShut {NoStop}%
\bibitem [{\citenamefont {Jouy}\ \emph {et~al.}(2011)\citenamefont {Jouy},
  \citenamefont {Vasanelli}, \citenamefont {Todorov}, \citenamefont {Delteil},
  \citenamefont {Biasiol}, \citenamefont {Sorba},\ and\ \citenamefont
  {Sirtori}}]{Jouy2011}%
  \BibitemOpen
  \bibfield  {author} {\bibinfo {author} {\bibfnamefont {P.}~\bibnamefont
  {Jouy}}, \bibinfo {author} {\bibfnamefont {A.}~\bibnamefont {Vasanelli}},
  \bibinfo {author} {\bibfnamefont {Y.}~\bibnamefont {Todorov}}, \bibinfo
  {author} {\bibfnamefont {A.}~\bibnamefont {Delteil}}, \bibinfo {author}
  {\bibfnamefont {G.}~\bibnamefont {Biasiol}}, \bibinfo {author} {\bibfnamefont
  {L.}~\bibnamefont {Sorba}}, \ and\ \bibinfo {author} {\bibfnamefont
  {C.}~\bibnamefont {Sirtori}},\ }\bibfield  {title} {\enquote {\bibinfo
  {title} {{Transition from strong to ultrastrong coupling regime in
  mid-infrared metal-dielectric-metal cavities}},}\ }\href {\doibase
  10.1063/1.3598432} {\bibfield  {journal} {\bibinfo  {journal} {Applied
  Physics Letters}\ }\textbf {\bibinfo {volume} {98}},\ \bibinfo {pages}
  {231114} (\bibinfo {year} {2011})},\ \Eprint {http://arxiv.org/abs/1212.4439}
  {arXiv:1212.4439} \BibitemShut {NoStop}%
\bibitem [{\citenamefont {Delteil}\ \emph {et~al.}(2012)\citenamefont
  {Delteil}, \citenamefont {Vasanelli}, \citenamefont {Todorov}, \citenamefont
  {{Feuillet Palma}}, \citenamefont {{Renaudat St-Jean}}, \citenamefont
  {Beaudoin}, \citenamefont {Sagnes},\ and\ \citenamefont
  {Sirtori}}]{Delteil2012}%
  \BibitemOpen
  \bibfield  {author} {\bibinfo {author} {\bibfnamefont {A.}~\bibnamefont
  {Delteil}}, \bibinfo {author} {\bibfnamefont {A.}~\bibnamefont {Vasanelli}},
  \bibinfo {author} {\bibfnamefont {Y.}~\bibnamefont {Todorov}}, \bibinfo
  {author} {\bibfnamefont {C.}~\bibnamefont {{Feuillet Palma}}}, \bibinfo
  {author} {\bibfnamefont {M.}~\bibnamefont {{Renaudat St-Jean}}}, \bibinfo
  {author} {\bibfnamefont {G.}~\bibnamefont {Beaudoin}}, \bibinfo {author}
  {\bibfnamefont {I.}~\bibnamefont {Sagnes}}, \ and\ \bibinfo {author}
  {\bibfnamefont {C.}~\bibnamefont {Sirtori}},\ }\bibfield  {title} {\enquote
  {\bibinfo {title} {{Charge-Induced Coherence between Intersubband Plasmons in
  a Quantum Structure}},}\ }\href {\doibase 10.1103/PhysRevLett.109.246808}
  {\bibfield  {journal} {\bibinfo  {journal} {Physical Review Letters}\
  }\textbf {\bibinfo {volume} {109}},\ \bibinfo {pages} {246808} (\bibinfo
  {year} {2012})},\ \Eprint {http://arxiv.org/abs/1212.4422} {arXiv:1212.4422}
  \BibitemShut {NoStop}%
\bibitem [{\citenamefont {Deng}\ \emph {et~al.}(2010)\citenamefont {Deng},
  \citenamefont {Haug},\ and\ \citenamefont {Yamamoto}}]{Deng2010}%
  \BibitemOpen
  \bibfield  {author} {\bibinfo {author} {\bibfnamefont {H.}~\bibnamefont
  {Deng}}, \bibinfo {author} {\bibfnamefont {H.}~\bibnamefont {Haug}}, \ and\
  \bibinfo {author} {\bibfnamefont {Y.}~\bibnamefont {Yamamoto}},\ }\bibfield
  {title} {\enquote {\bibinfo {title} {{Exciton-polariton Bose-Einstein
  condensation}},}\ }\href {\doibase 10.1103/RevModPhys.82.1489} {\bibfield
  {journal} {\bibinfo  {journal} {Reviews of Modern Physics}\ }\textbf
  {\bibinfo {volume} {82}},\ \bibinfo {pages} {1489} (\bibinfo {year}
  {2010})}\BibitemShut {NoStop}%
\bibitem [{\citenamefont {Coropceanu}\ \emph {et~al.}(2007)\citenamefont
  {Coropceanu}, \citenamefont {Cornil}, \citenamefont {{da Silva Filho}},
  \citenamefont {Olivier}, \citenamefont {Silbey},\ and\ \citenamefont
  {Br{\'{e}}das}}]{Coropceanu2007}%
  \BibitemOpen
  \bibfield  {author} {\bibinfo {author} {\bibfnamefont {V.}~\bibnamefont
  {Coropceanu}}, \bibinfo {author} {\bibfnamefont {J.}~\bibnamefont {Cornil}},
  \bibinfo {author} {\bibfnamefont {D.~A.}\ \bibnamefont {{da Silva Filho}}},
  \bibinfo {author} {\bibfnamefont {Y.}~\bibnamefont {Olivier}}, \bibinfo
  {author} {\bibfnamefont {R.}~\bibnamefont {Silbey}}, \ and\ \bibinfo {author}
  {\bibfnamefont {J.-L.}\ \bibnamefont {Br{\'{e}}das}},\ }\bibfield  {title}
  {\enquote {\bibinfo {title} {{Charge Transport in Organic Semiconductors}},}\
  }\href {\doibase 10.1021/cr050140x} {\bibfield  {journal} {\bibinfo
  {journal} {Chemical Reviews}\ }\textbf {\bibinfo {volume} {107}},\ \bibinfo
  {pages} {926} (\bibinfo {year} {2007})}\BibitemShut {NoStop}%
\bibitem [{\citenamefont {Nagarajan}\ \emph {et~al.}(2020)\citenamefont
  {Nagarajan}, \citenamefont {George}, \citenamefont {Thomas}, \citenamefont
  {Devaux}, \citenamefont {Chervy}, \citenamefont {Azzini}, \citenamefont
  {Joseph}, \citenamefont {Jouaiti}, \citenamefont {Hosseini}, \citenamefont
  {Kumar}, \citenamefont {Genet}, \citenamefont {Bartolo}, \citenamefont
  {Ciuti},\ and\ \citenamefont {Ebbesen}}]{Nagarajan2020}%
  \BibitemOpen
  \bibfield  {author} {\bibinfo {author} {\bibfnamefont {K.}~\bibnamefont
  {Nagarajan}}, \bibinfo {author} {\bibfnamefont {J.}~\bibnamefont {George}},
  \bibinfo {author} {\bibfnamefont {A.}~\bibnamefont {Thomas}}, \bibinfo
  {author} {\bibfnamefont {E.}~\bibnamefont {Devaux}}, \bibinfo {author}
  {\bibfnamefont {T.}~\bibnamefont {Chervy}}, \bibinfo {author} {\bibfnamefont
  {S.}~\bibnamefont {Azzini}}, \bibinfo {author} {\bibfnamefont
  {K.}~\bibnamefont {Joseph}}, \bibinfo {author} {\bibfnamefont
  {A.}~\bibnamefont {Jouaiti}}, \bibinfo {author} {\bibfnamefont {M.~W.}\
  \bibnamefont {Hosseini}}, \bibinfo {author} {\bibfnamefont {A.}~\bibnamefont
  {Kumar}}, \bibinfo {author} {\bibfnamefont {C.}~\bibnamefont {Genet}},
  \bibinfo {author} {\bibfnamefont {N.}~\bibnamefont {Bartolo}}, \bibinfo
  {author} {\bibfnamefont {C.}~\bibnamefont {Ciuti}}, \ and\ \bibinfo {author}
  {\bibfnamefont {T.~W.}\ \bibnamefont {Ebbesen}},\ }\bibfield  {title}
  {\enquote {\bibinfo {title} {{Conductivity and Photoconductivity of a p-Type
  Organic Semiconductor under Ultrastrong Coupling}},}\ }\href {\doibase
  10.1021/acsnano.0c03496} {\bibfield  {journal} {\bibinfo  {journal} {ACS
  Nano}\ }\textbf {\bibinfo {volume} {14}},\ \bibinfo {pages} {10219} (\bibinfo
  {year} {2020})}\BibitemShut {NoStop}%
\bibitem [{\citenamefont {Orgiu}\ \emph {et~al.}(2015)\citenamefont {Orgiu},
  \citenamefont {George}, \citenamefont {Hutchison}, \citenamefont {Devaux},
  \citenamefont {Dayen}, \citenamefont {Doudin}, \citenamefont {Stellacci},
  \citenamefont {Genet}, \citenamefont {Schachenmayer}, \citenamefont {Genes},
  \citenamefont {Pupillo}, \citenamefont {Samor{\`{i}}},\ and\ \citenamefont
  {Ebbesen}}]{Orgiu2015}%
  \BibitemOpen
  \bibfield  {author} {\bibinfo {author} {\bibfnamefont {E.}~\bibnamefont
  {Orgiu}}, \bibinfo {author} {\bibfnamefont {J.}~\bibnamefont {George}},
  \bibinfo {author} {\bibfnamefont {J.~A.}\ \bibnamefont {Hutchison}}, \bibinfo
  {author} {\bibfnamefont {E.}~\bibnamefont {Devaux}}, \bibinfo {author}
  {\bibfnamefont {J.~F.}\ \bibnamefont {Dayen}}, \bibinfo {author}
  {\bibfnamefont {B.}~\bibnamefont {Doudin}}, \bibinfo {author} {\bibfnamefont
  {F.}~\bibnamefont {Stellacci}}, \bibinfo {author} {\bibfnamefont
  {C.}~\bibnamefont {Genet}}, \bibinfo {author} {\bibfnamefont
  {J.}~\bibnamefont {Schachenmayer}}, \bibinfo {author} {\bibfnamefont
  {C.}~\bibnamefont {Genes}}, \bibinfo {author} {\bibfnamefont
  {G.}~\bibnamefont {Pupillo}}, \bibinfo {author} {\bibfnamefont
  {P.}~\bibnamefont {Samor{\`{i}}}}, \ and\ \bibinfo {author} {\bibfnamefont
  {T.~W.}\ \bibnamefont {Ebbesen}},\ }\bibfield  {title} {\enquote {\bibinfo
  {title} {{Conductivity in organic semiconductors hybridized with the vacuum
  field}},}\ }\href {\doibase 10.1038/nmat4392} {\bibfield  {journal} {\bibinfo
   {journal} {Nature Materials}\ }\textbf {\bibinfo {volume} {14}},\ \bibinfo
  {pages} {1123} (\bibinfo {year} {2015})},\ \Eprint
  {http://arxiv.org/abs/1409.1900} {arXiv:1409.1900} \BibitemShut {NoStop}%
\bibitem [{\citenamefont {Kang}\ \emph {et~al.}(2021)\citenamefont {Kang},
  \citenamefont {Chen}, \citenamefont {Derek}, \citenamefont {H{\"{a}}gglund},
  \citenamefont {G{\l}owacki},\ and\ \citenamefont {Jonsson}}]{Kang2021}%
  \BibitemOpen
  \bibfield  {author} {\bibinfo {author} {\bibfnamefont {E.~S.~H.}\
  \bibnamefont {Kang}}, \bibinfo {author} {\bibfnamefont {S.}~\bibnamefont
  {Chen}}, \bibinfo {author} {\bibfnamefont {V.}~\bibnamefont {Derek}},
  \bibinfo {author} {\bibfnamefont {C.}~\bibnamefont {H{\"{a}}gglund}},
  \bibinfo {author} {\bibfnamefont {E.~D.}\ \bibnamefont {G{\l}owacki}}, \ and\
  \bibinfo {author} {\bibfnamefont {M.~P.}\ \bibnamefont {Jonsson}},\
  }\bibfield  {title} {\enquote {\bibinfo {title} {{Charge transport in
  phthalocyanine thin-film transistors coupled with Fabry-Perot cavities}},}\
  }\href {\doibase 10.1039/D0TC05418F} {\bibfield  {journal} {\bibinfo
  {journal} {Journal of Materials Chemistry C}\ }\textbf {\bibinfo {volume}
  {9}},\ \bibinfo {pages} {2368} (\bibinfo {year} {2021})}\BibitemShut
  {NoStop}%
\bibitem [{\citenamefont {Bhatt}\ \emph {et~al.}(2021)\citenamefont {Bhatt},
  \citenamefont {Kaur},\ and\ \citenamefont {George}}]{Bhatt2021}%
  \BibitemOpen
  \bibfield  {author} {\bibinfo {author} {\bibfnamefont {P.}~\bibnamefont
  {Bhatt}}, \bibinfo {author} {\bibfnamefont {K.}~\bibnamefont {Kaur}}, \ and\
  \bibinfo {author} {\bibfnamefont {J.}~\bibnamefont {George}},\ }\bibfield
  {title} {\enquote {\bibinfo {title} {{Enhanced Charge Transport in
  Two-Dimensional Materials through Light–Matter Strong Coupling}},}\ }\href
  {\doibase 10.1021/acsnano.1c04544} {\bibfield  {journal} {\bibinfo  {journal}
  {ACS Nano}\ }\textbf {\bibinfo {volume} {15}},\ \bibinfo {pages} {13616}
  (\bibinfo {year} {2021})}\BibitemShut {NoStop}%
\bibitem [{\citenamefont {Moore}(1970)}]{Moore1970}%
  \BibitemOpen
  \bibfield  {author} {\bibinfo {author} {\bibfnamefont {G.~T.}\ \bibnamefont
  {Moore}},\ }\bibfield  {title} {\enquote {\bibinfo {title} {{Quantum Theory
  of the Electromagnetic Field in a Variable-Length One-Dimensional Cavity}},}\
  }\href {\doibase 10.1063/1.1665432} {\bibfield  {journal} {\bibinfo
  {journal} {Journal of Mathematical Physics}\ }\textbf {\bibinfo {volume}
  {11}},\ \bibinfo {pages} {2679} (\bibinfo {year} {1970})}\BibitemShut
  {NoStop}%
\bibitem [{\citenamefont {Cirio}\ \emph {et~al.}(2019)\citenamefont {Cirio},
  \citenamefont {Shammah}, \citenamefont {Lambert}, \citenamefont {{De
  Liberato}},\ and\ \citenamefont {Nori}}]{Cirio2019}%
  \BibitemOpen
  \bibfield  {author} {\bibinfo {author} {\bibfnamefont {M.}~\bibnamefont
  {Cirio}}, \bibinfo {author} {\bibfnamefont {N.}~\bibnamefont {Shammah}},
  \bibinfo {author} {\bibfnamefont {N.}~\bibnamefont {Lambert}}, \bibinfo
  {author} {\bibfnamefont {S.}~\bibnamefont {{De Liberato}}}, \ and\ \bibinfo
  {author} {\bibfnamefont {F.}~\bibnamefont {Nori}},\ }\bibfield  {title}
  {\enquote {\bibinfo {title} {{Multielectron Ground State
  Electroluminescence}},}\ }\href {\doibase 10.1103/PhysRevLett.122.190403}
  {\bibfield  {journal} {\bibinfo  {journal} {Physical Review Letters}\
  }\textbf {\bibinfo {volume} {122}},\ \bibinfo {pages} {190403} (\bibinfo
  {year} {2019})},\ \Eprint {http://arxiv.org/abs/1811.08682}
  {arXiv:1811.08682} \BibitemShut {NoStop}%
\end{thebibliography}%


\begin{thebibliography}{2}%
\makeatletter
\providecommand \@ifxundefined [1]{%
 \@ifx{#1\undefined}
}%
\providecommand \@ifnum [1]{%
 \ifnum #1\expandafter \@firstoftwo
 \else \expandafter \@secondoftwo
 \fi
}%
\providecommand \@ifx [1]{%
 \ifx #1\expandafter \@firstoftwo
 \else \expandafter \@secondoftwo
 \fi
}%
\providecommand \natexlab [1]{#1}%
\providecommand \enquote  [1]{``#1''}%
\providecommand \bibnamefont  [1]{#1}%
\providecommand \bibfnamefont [1]{#1}%
\providecommand \citenamefont [1]{#1}%
\providecommand \href@noop [0]{\@secondoftwo}%
\providecommand \href [0]{\begingroup \@sanitize@url \@href}%
\providecommand \@href[1]{\@@startlink{#1}\@@href}%
\providecommand \@@href[1]{\endgroup#1\@@endlink}%
\providecommand \@sanitize@url [0]{\catcode `\\12\catcode `\$12\catcode
  `\&12\catcode `\#12\catcode `\^12\catcode `\_12\catcode `\%12\relax}%
\providecommand \@@startlink[1]{}%
\providecommand \@@endlink[0]{}%
\providecommand \url  [0]{\begingroup\@sanitize@url \@url }%
\providecommand \@url [1]{\endgroup\@href {#1}{\urlprefix }}%
\providecommand \urlprefix  [0]{URL }%
\providecommand \Eprint [0]{\href }%
\providecommand \doibase [0]{http://dx.doi.org/}%
\providecommand \selectlanguage [0]{\@gobble}%
\providecommand \bibinfo  [0]{\@secondoftwo}%
\providecommand \bibfield  [0]{\@secondoftwo}%
\providecommand \translation [1]{[#1]}%
\providecommand \BibitemOpen [0]{}%
\providecommand \bibitemStop [0]{}%
\providecommand \bibitemNoStop [0]{.\EOS\space}%
\providecommand \EOS [0]{\spacefactor3000\relax}%
\providecommand \BibitemShut  [1]{\csname bibitem#1\endcsname}%
\let\auto@bib@innerbib\@empty
\bibitem [{\citenamefont {Nishiuchi}\ \emph {et~al.}(2018)\citenamefont
  {Nishiuchi}, \citenamefont {Aibara},\ and\ \citenamefont
  {Kubo}}]{Nishiuchi2018}%
  \BibitemOpen
  \bibfield  {author} {\bibinfo {author} {\bibfnamefont {T.}~\bibnamefont
  {Nishiuchi}}, \bibinfo {author} {\bibfnamefont {S.}~\bibnamefont {Aibara}}, \
  and\ \bibinfo {author} {\bibfnamefont {T.}~\bibnamefont {Kubo}},\ }\bibfield
  {title} {\enquote {\bibinfo {title} {{Synthesis and Properties of a Highly
  Congested Tri(9-anthryl)methyl Radical}},}\ }\href {\doibase
  10.1002/anie.201811314} {\bibfield  {journal} {\bibinfo  {journal}
  {Angewandte Chemie International Edition}\ }\textbf {\bibinfo {volume}
  {57}},\ \bibinfo {pages} {16516} (\bibinfo {year} {2018})}\BibitemShut
  {NoStop}%
\bibitem [{\citenamefont {Frisch}\ \emph {et~al.}(2016)\citenamefont {Frisch},
  \citenamefont {Trucks}, \citenamefont {Schlegel}, \citenamefont {Scuseria},
  \citenamefont {Robb}, \citenamefont {Cheeseman}, \citenamefont {Scalmani},
  \citenamefont {Barone}, \citenamefont {Petersson}, \citenamefont {Nakatsuji},
  \citenamefont {Li}, \citenamefont {Caricato}, \citenamefont {Marenich},
  \citenamefont {Bloino}, \citenamefont {Janesko}, \citenamefont {Gomperts},
  \citenamefont {Mennucci}, \citenamefont {Hratchian}, \citenamefont {Ortiz},
  \citenamefont {Izmaylov}, \citenamefont {Sonnenberg}, \citenamefont
  {Williams-Young}, \citenamefont {Ding}, \citenamefont {Lipparini},
  \citenamefont {Egidi}, \citenamefont {Goings}, \citenamefont {Peng},
  \citenamefont {Petrone}, \citenamefont {Henderson}, \citenamefont
  {Ranasinghe}, \citenamefont {Zakrzewski}, \citenamefont {Gao}, \citenamefont
  {Rega}, \citenamefont {Zheng}, \citenamefont {Liang}, \citenamefont {Hada},
  \citenamefont {Ehara}, \citenamefont {Toyota}, \citenamefont {Fukuda},
  \citenamefont {Hasegawa}, \citenamefont {Ishida}, \citenamefont {Nakajima},
  \citenamefont {Honda}, \citenamefont {Kitao}, \citenamefont {Nakai},
  \citenamefont {Vreven}, \citenamefont {Throssell}, \citenamefont
  {Montgomery}, \citenamefont {Peralta}, \citenamefont {Ogliaro}, \citenamefont
  {Bearpark}, \citenamefont {Heyd}, \citenamefont {Brothers}, \citenamefont
  {Kudin}, \citenamefont {Staroverov}, \citenamefont {Keith}, \citenamefont
  {Kobayashi}, \citenamefont {Normand}, \citenamefont {Raghavachari},
  \citenamefont {Rendell}, \citenamefont {Burant}, \citenamefont {Iyengar},
  \citenamefont {Tomasi}, \citenamefont {Cossi}, \citenamefont {Millam},
  \citenamefont {Klene}, \citenamefont {Adamo}, \citenamefont {Cammi},
  \citenamefont {Ochterski}, \citenamefont {Martin}, \citenamefont {Morokuma},
  \citenamefont {Farkas}, \citenamefont {Foresman},\ and\ \citenamefont
  {Fox}}]{Frisch2016}%
  \BibitemOpen
  \bibfield  {author} {\bibinfo {author} {\bibfnamefont {M.~J.}\ \bibnamefont
  {Frisch}}, \bibinfo {author} {\bibfnamefont {G.~W.}\ \bibnamefont {Trucks}},
  \bibinfo {author} {\bibfnamefont {H.~B.}\ \bibnamefont {Schlegel}}, \bibinfo
  {author} {\bibfnamefont {G.~E.}\ \bibnamefont {Scuseria}}, \bibinfo {author}
  {\bibfnamefont {M.~A.}\ \bibnamefont {Robb}}, \bibinfo {author}
  {\bibfnamefont {J.~R.}\ \bibnamefont {Cheeseman}}, \bibinfo {author}
  {\bibfnamefont {G.}~\bibnamefont {Scalmani}}, \bibinfo {author}
  {\bibfnamefont {V.}~\bibnamefont {Barone}}, \bibinfo {author} {\bibfnamefont
  {G.~A.}\ \bibnamefont {Petersson}}, \bibinfo {author} {\bibfnamefont
  {H.}~\bibnamefont {Nakatsuji}}, \bibinfo {author} {\bibfnamefont
  {X.}~\bibnamefont {Li}}, \bibinfo {author} {\bibfnamefont {M.}~\bibnamefont
  {Caricato}}, \bibinfo {author} {\bibfnamefont {A.~V.}\ \bibnamefont
  {Marenich}}, \bibinfo {author} {\bibfnamefont {J.}~\bibnamefont {Bloino}},
  \bibinfo {author} {\bibfnamefont {B.~G.}\ \bibnamefont {Janesko}}, \bibinfo
  {author} {\bibfnamefont {R.}~\bibnamefont {Gomperts}}, \bibinfo {author}
  {\bibfnamefont {B.}~\bibnamefont {Mennucci}}, \bibinfo {author}
  {\bibfnamefont {H.~P.}\ \bibnamefont {Hratchian}}, \bibinfo {author}
  {\bibfnamefont {J.~V.}\ \bibnamefont {Ortiz}}, \bibinfo {author}
  {\bibfnamefont {A.~F.}\ \bibnamefont {Izmaylov}}, \bibinfo {author}
  {\bibfnamefont {J.~L.}\ \bibnamefont {Sonnenberg}}, \bibinfo {author}
  {\bibfnamefont {D.}~\bibnamefont {Williams-Young}}, \bibinfo {author}
  {\bibfnamefont {F.}~\bibnamefont {Ding}}, \bibinfo {author} {\bibfnamefont
  {F.}~\bibnamefont {Lipparini}}, \bibinfo {author} {\bibfnamefont
  {F.}~\bibnamefont {Egidi}}, \bibinfo {author} {\bibfnamefont
  {J.}~\bibnamefont {Goings}}, \bibinfo {author} {\bibfnamefont
  {B.}~\bibnamefont {Peng}}, \bibinfo {author} {\bibfnamefont {A.}~\bibnamefont
  {Petrone}}, \bibinfo {author} {\bibfnamefont {T.}~\bibnamefont {Henderson}},
  \bibinfo {author} {\bibfnamefont {D.}~\bibnamefont {Ranasinghe}}, \bibinfo
  {author} {\bibfnamefont {V.~G.}\ \bibnamefont {Zakrzewski}}, \bibinfo
  {author} {\bibfnamefont {J.}~\bibnamefont {Gao}}, \bibinfo {author}
  {\bibfnamefont {N.}~\bibnamefont {Rega}}, \bibinfo {author} {\bibfnamefont
  {G.}~\bibnamefont {Zheng}}, \bibinfo {author} {\bibfnamefont
  {W.}~\bibnamefont {Liang}}, \bibinfo {author} {\bibfnamefont
  {M.}~\bibnamefont {Hada}}, \bibinfo {author} {\bibfnamefont {M.}~\bibnamefont
  {Ehara}}, \bibinfo {author} {\bibfnamefont {K.}~\bibnamefont {Toyota}},
  \bibinfo {author} {\bibfnamefont {R.}~\bibnamefont {Fukuda}}, \bibinfo
  {author} {\bibfnamefont {J.}~\bibnamefont {Hasegawa}}, \bibinfo {author}
  {\bibfnamefont {M.}~\bibnamefont {Ishida}}, \bibinfo {author} {\bibfnamefont
  {T.}~\bibnamefont {Nakajima}}, \bibinfo {author} {\bibfnamefont
  {Y.}~\bibnamefont {Honda}}, \bibinfo {author} {\bibfnamefont
  {O.}~\bibnamefont {Kitao}}, \bibinfo {author} {\bibfnamefont
  {H.}~\bibnamefont {Nakai}}, \bibinfo {author} {\bibfnamefont
  {T.}~\bibnamefont {Vreven}}, \bibinfo {author} {\bibfnamefont
  {K.}~\bibnamefont {Throssell}}, \bibinfo {author} {\bibfnamefont {J.~A.}\
  \bibnamefont {Montgomery}, \bibfnamefont {{Jr.}}}, \bibinfo {author}
  {\bibfnamefont {J.~E.}\ \bibnamefont {Peralta}}, \bibinfo {author}
  {\bibfnamefont {F.}~\bibnamefont {Ogliaro}}, \bibinfo {author} {\bibfnamefont
  {M.~J.}\ \bibnamefont {Bearpark}}, \bibinfo {author} {\bibfnamefont {J.~J.}\
  \bibnamefont {Heyd}}, \bibinfo {author} {\bibfnamefont {E.~N.}\ \bibnamefont
  {Brothers}}, \bibinfo {author} {\bibfnamefont {K.~N.}\ \bibnamefont {Kudin}},
  \bibinfo {author} {\bibfnamefont {V.~N.}\ \bibnamefont {Staroverov}},
  \bibinfo {author} {\bibfnamefont {T.~A.}\ \bibnamefont {Keith}}, \bibinfo
  {author} {\bibfnamefont {R.}~\bibnamefont {Kobayashi}}, \bibinfo {author}
  {\bibfnamefont {J.}~\bibnamefont {Normand}}, \bibinfo {author} {\bibfnamefont
  {K.}~\bibnamefont {Raghavachari}}, \bibinfo {author} {\bibfnamefont {A.~P.}\
  \bibnamefont {Rendell}}, \bibinfo {author} {\bibfnamefont {J.~C.}\
  \bibnamefont {Burant}}, \bibinfo {author} {\bibfnamefont {S.~S.}\
  \bibnamefont {Iyengar}}, \bibinfo {author} {\bibfnamefont {J.}~\bibnamefont
  {Tomasi}}, \bibinfo {author} {\bibfnamefont {M.}~\bibnamefont {Cossi}},
  \bibinfo {author} {\bibfnamefont {J.~M.}\ \bibnamefont {Millam}}, \bibinfo
  {author} {\bibfnamefont {M.}~\bibnamefont {Klene}}, \bibinfo {author}
  {\bibfnamefont {C.}~\bibnamefont {Adamo}}, \bibinfo {author} {\bibfnamefont
  {R.}~\bibnamefont {Cammi}}, \bibinfo {author} {\bibfnamefont {J.~W.}\
  \bibnamefont {Ochterski}}, \bibinfo {author} {\bibfnamefont {R.~L.}\
  \bibnamefont {Martin}}, \bibinfo {author} {\bibfnamefont {K.}~\bibnamefont
  {Morokuma}}, \bibinfo {author} {\bibfnamefont {O.}~\bibnamefont {Farkas}},
  \bibinfo {author} {\bibfnamefont {J.~B.}\ \bibnamefont {Foresman}}, \ and\
  \bibinfo {author} {\bibfnamefont {D.~J.}\ \bibnamefont {Fox}},\ }\href@noop
  {} {\enquote {\bibinfo {title} {Gaussian 16 {R}evision {C}.01},}\ } (\bibinfo
  {year} {2016}),\ \bibinfo {note} {gaussian Inc. Wallingford CT}\BibitemShut
  {NoStop}%
\end{thebibliography}%

\end{document}


\title{Supplemental Material for ``Organic charged polaritons in the ultrastrong coupling regime''}

\author{Mao Wang}
\affiliation{Department of Chemistry and Molecular Biology, University of Gothenburg, 412 96 Gothenburg, Sweden}

\author{Suman Mallick}
\affiliation{Department of Chemistry and Molecular Biology, University of Gothenburg, 412 96 Gothenburg, Sweden}

\author{Anton Frisk Kockum}
\affiliation{Department of Microtechnology and Nanoscience, Chalmers University of Technology, 412 96 Gothenburg, Sweden}

\author{Karl B\"orjesson}
\affiliation{Department of Chemistry and Molecular Biology, University of Gothenburg, 412 96 Gothenburg, Sweden}

\date{\today}


\maketitle

\tableofcontents

\renewcommand{\thefigure}{S\arabic{figure}}
\renewcommand{\thesection}{S\arabic{section}}
\renewcommand{\theequation}{S\arabic{equation}}
\renewcommand{\thetable}{S\arabic{table}}
\renewcommand{\bibnumfmt}[1]{[S#1]}
\renewcommand{\citenumfont}[1]{S#1}


\section{Synthesis}


\subsection{General}

All starting materials were purchased from Sigma-Aldrich Chemical Co.~and used without further purification. All moisture- and oxygen-sensitive reactions were carried out using Schlenk techniques in oven-dried glassware. Solvents used for moisture and oxygen-sensitive reactions were dried using an MBraun MB SPS-800 solvent purification system and, if necessary, degassed by freeze-pump-thaw cycles and stored over \unit[4]{\AA} molecular sieves under argon atmosphere. Flash chromatography was performed using a Teledyne CombiFlash EZ prep and normal-phase silica. $^1$H nuclear magnetic resonance (NMR) spectra were recorded on a Varian spectrometer at \unit[400]{MHz}, J-coupling values are given in Hertz, and chemical shifts are given in parts per million using tetramethysilane, with \unit[0.00]{ppm} as an internal standard.

\begin{figure}[]
\includegraphics[width=\linewidth]{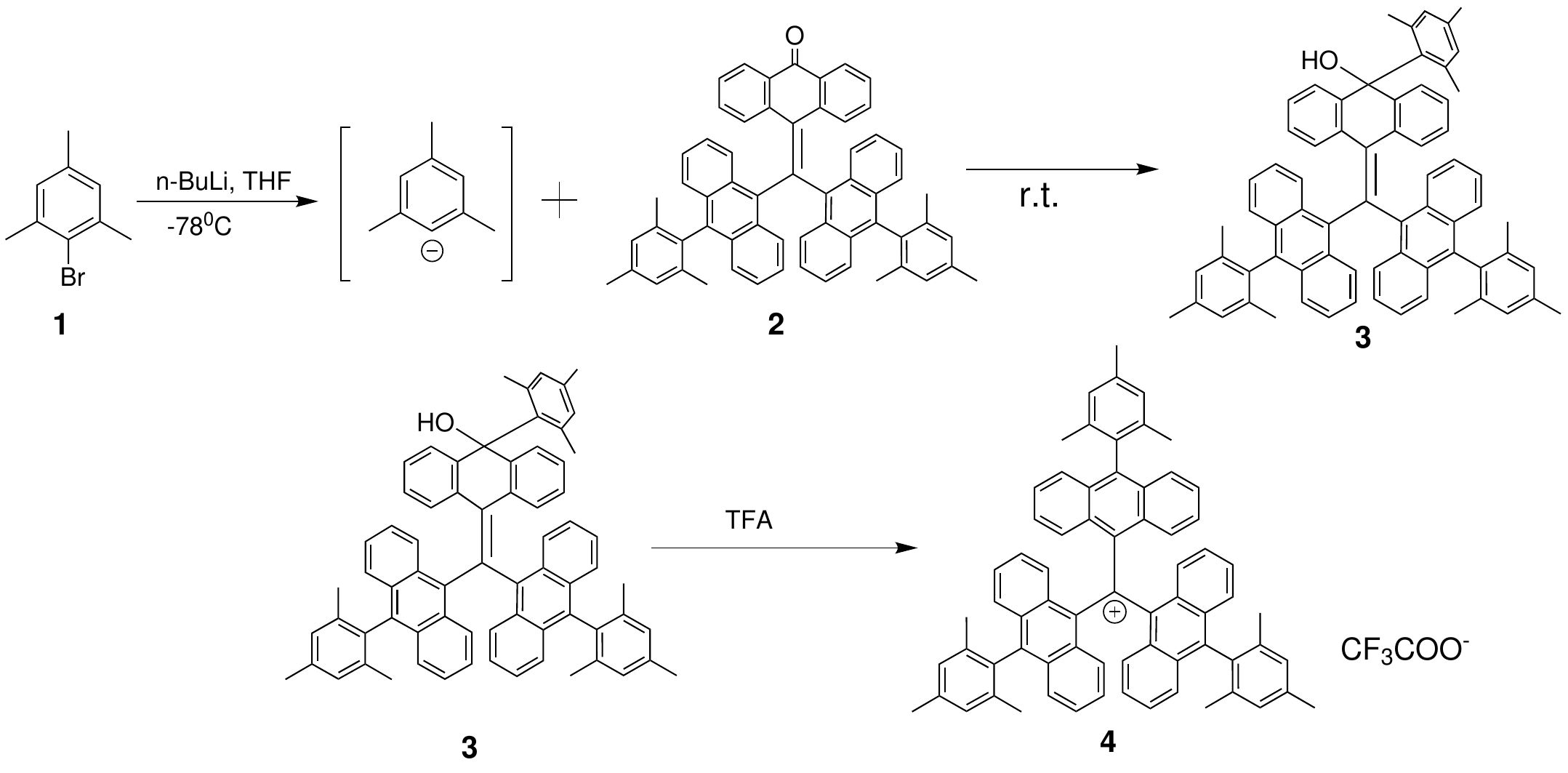}
\caption{
Synthesis scheme.
\label{fig:SynthesisScheme}}
\end{figure}

The full synthesis scheme is shown in \figref{fig:SynthesisScheme}.


\subsection{Synthesis of compound 3}

\begin{figure}[]
\includegraphics[width=\linewidth]{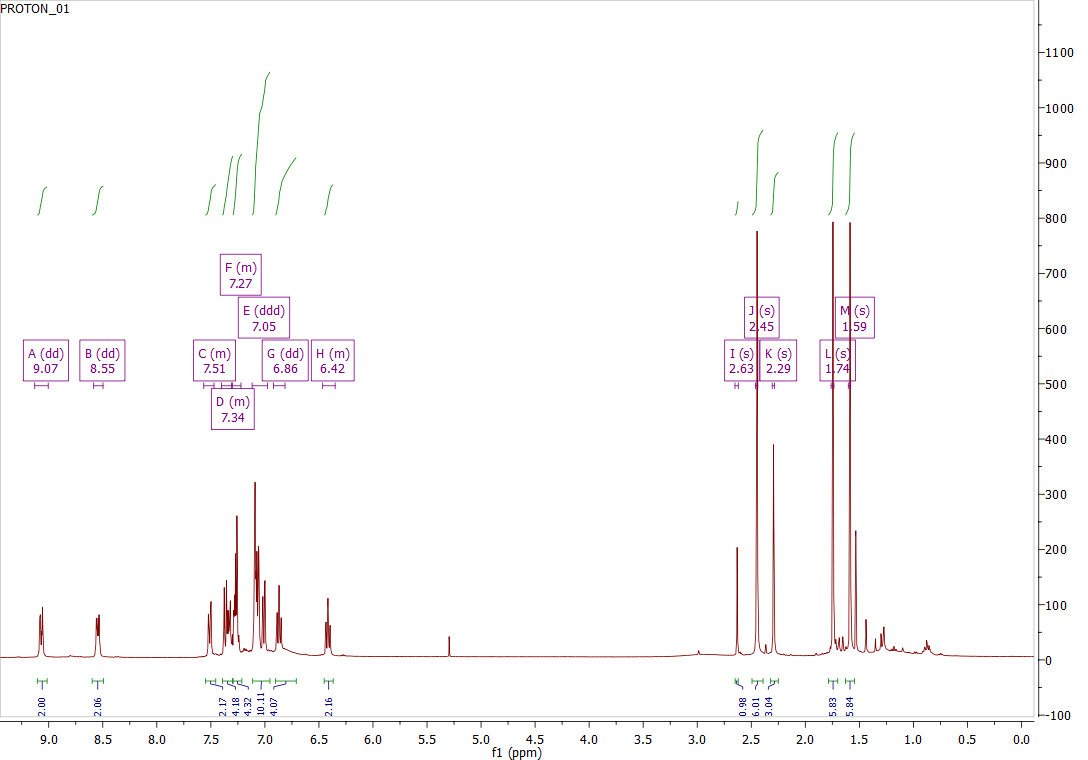}
\caption{
$^1$H NMR of the alcohol precursor used in the synthesis. See the text for extracted values.
\label{fig:NMR3}}
\end{figure}


2-bromomesitylene (\unit[425]{mg}, \unit[2.12]{mmol}) was dissolved in dry THF (\unit[15]{ml}), cooled at \unit[$-78$]{$^\circ$C}, then $n$-BuLi (\unit[1.6]{M} hexane solution, \unit[1.3]{ml}, \unit[0.62]{mmol}) was added dropwise to the stirring solution. Compound \textbf{2} was prepared according to the literature procedure beforehand~\cite{Nishiuchi2018}. After stirring for \unit[40]{min} at the same temperature, a solution of compound \textbf{2} (\unit[425]{mg}, \unit[0.53]{mmol}) in dry THF (\unit[15]{ml}) was added and stirred for \unit[3.5]{h} at room temperature. The reaction was monitored using TLC and finally quenched by adding \unit[5]{ml} water. The reaction mixture was extracted with dichloromethane ($\unit[2 \times 50]{ml}$), then washed with brine and dried over sodium sulfate. After removal of the solvent in vacuo, the crude material was purified by column chromatography on silica gel using dichloromethane and hexane as eluent to afford the compound \textbf{3} (\unit[244]{mg}, \unit[0.26]{mmol}, \unit[53]{\%}) as deep orange solid. NMR (see \figref{fig:NMR3} and values below) was in accordance with the literature~\cite{Nishiuchi2018}.

$^1$H NMR (\unit[400]{MHz}, CDCl3): $\delta$ 9.07 (dd, 2H, aromatic proton), 8.5d (dd, 2H, aromatic proton), 7.51--7.48 (m, 2H, aromatic proton), 7.36--7.29 (m, 4H, aromatic proton), 7.27 (m, 4H, aromatic proton), 7.09--6.98 (m, 10H, aromatic proton), 6.90--6.60 (t + br, $J = \unit[7.6]{Hz}$ for t, 4H, aromatic proton), 6.42 (t, $J = \unit[8.4]{Hz}$, 2H, aromatic proton), 2.62 (s, 1H, OH), 2.44 (s, 6H), 2.28 (s, 3H), 1.73 (s, 6H), 1.57 (s, 6H).


\subsection{Synthesis of tris(10-mesitylanthr-9-yl)methyl cation (4)}

\begin{figure}[]
\includegraphics[width=\linewidth]{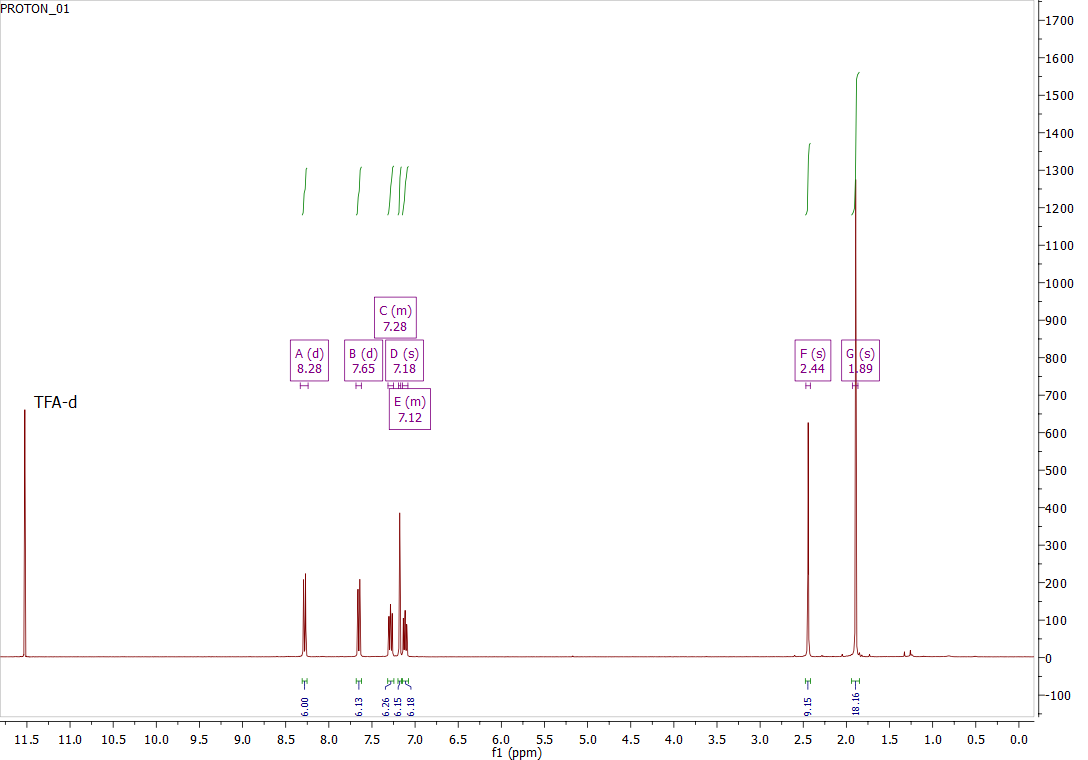}
\caption{
$^1$H NMR of the of the main product. See the text for extracted values.
\label{fig:NMR4}}
\end{figure}


Compound \textbf{3} (\unit[30]{mg}) obtained in the previous step was dissolved in \unit[1]{ml} trifluoroacetic acid (TFA) and the solution was stirred for 2 minutes. The solution turned to deep red in color and the yield was quantitative, as evidenced by NMR (see \figref{fig:NMR4}), using TFA-d as the solvent.

$^1$H NMR (\unit[400]{MHz}, TFA-d): $\delta$ 8.28 (d, $J = \unit[9.2]{Hz}$, 6H), 7.65 (d, $J = \unit[8.8]{Hz}$, 6H), 7.28 (t, $J = \unit[7.4]{Hz}$, 6H), 7.18 (s, 6H), 7.12 (m, 6H), 2.44 (s, 9H), 1.89 (s, 18H).


\section{Cavity fabrication and characterization}

Au mirrors of the Fabry-P\'erot cavities were fabricated by vacuum sputtering deposition (HEX, Korvus Technologies). A semitransparent Au ($\sim\,$\unit[22]{nm}) film was sputtered on top of a precleaned glass substrate as the bottom mirror. Compound \textbf{3} (\unit[3.85]{mg}) was dissolved in trifluoroacetic acid ($\unit[100]{\mu L}$) to form the TAntM cation solution before spin-coating (\unit[1500]{rpm}) the film on the Au-covered substrates. A top Au mirror ($\sim\,$\unit[22]{nm}) was sputtered on the film to complete the cavity. Steady-state absorption spectra were measured using a standard spectrophotometer (Lambda 950, Perkin Elmer). The angle-resolved transmission spectra were measured using the same spectrophotometer equipped with a rotatable sample holder.


\section{Quantum-mechanical calculations}

The Mulliken atomic charges were calculated using density functional theory using the Gaussian 16 program package~\cite{Frisch2016}. A ground-state energy calculation was performed using the geometry of the published crystal structure~\cite{Nishiuchi2018}, the functional m06, and the basis set 6-311G+.


\bibliography{CationUSCRefs}